\documentclass[pra,onecolumn,showpacs]{revtex4-1}
\usepackage{amssymb}
\usepackage{amsmath}
\usepackage{graphicx}
\usepackage{subfigure}
\usepackage{epstopdf}
\usepackage{color}

\begin{document}

\title{Spontaneous symmetry breaking in a split potential box}
\author{Elad Shamriz$^{1}$, Nir Dror$^{1}$, and Boris A. Malomed$^{1,2}$}
\affiliation{$^1$Department of Physical Electronics, School of Electrical Engineering,
Faculty of Engineering, Tel Aviv University, Tel Aviv 69978, Israel\\
$^2$Laboratory of Nonlinear-Optical Informatics, ITMO University, St. Petersburg 197101, Russia}

\begin{abstract}
We report results of the analysis of the spontaneous symmetry breaking (SSB)
in the basic (actually, simplest) model which is capable to produce the SSB
phenomenology in the one-dimensional setting. It is based on the
Gross-Pitaevskii -- nonlinear Schr\"{o}dinger equation with the cubic
self-attractive term and a double-well-potential built as an infinitely deep
potential box split by a narrow (delta-functional) barrier. The barrier's
strength, $\varepsilon $, is the single free parameter of the scaled form of
the model. It may be implemented in atomic Bose-Einstein condensates and
nonlinear optics. The SSB bifurcation of the symmetric ground state (GS) is
predicted analytically in two limit cases, \textit{viz}, for deep or weak
splitting of the potential box by the barrier ($\varepsilon \gg 1$ or $%
\varepsilon \ll 1$, respectively). 
For the generic case, a variational approximation (VA) is elaborated. The
analytical findings are presented along with systematic numerical results.
Stability of stationary states is studied through the calculation of
eigenvalues for small perturbations, and by means of direct simulations. The
GS always undergoes the SSB bifurcation of the supercritical type, as
predicted by the VA at moderate values of $\varepsilon $, although the VA
fails at small $\varepsilon $, due to inapplicability of the underlying
\textit{ansatz }in that case. However, the latter case is correctly treated
by the approximation based on a soliton ansatz. On top of the GS, the first
and second excited states are studied too. The antisymmetric mode (the first
excited state) is destabilized at a critical value of its norm. The second
excited state undergoes the SSB bifurcation, like the GS, but, unlike it,
the bifurcation produces an unstable asymmetric mode. All unstable modes
tend to spontaneously reshape into the asymmetric GS.
\end{abstract}

\maketitle

\section{Introduction and the model}

Dynamics of confined collective excitations in nonlinear physical systems,
modeled by one or several fields, is determined by the interplay of the
field-trapping potential and the character of interactions of the field(s).
In particular, the system's spatial symmetry is determined by the shape of
the potential. A generic type of the latter is represented by double-well
potentials (DWPs), which feature symmetry between two wells separated by a
barrier.

In quantum mechanics \cite{LL} and linear field theories mathematically
similar to it, such as paraxial light propagation in linear optical
waveguides \cite{SnyderLove}, the ground state (GS) of a confined system
normally follows the symmetry of the trapping potential (the Jahn-Teller
effect in molecules exemplifies another possibility, when the GS of the
electron wave function in the complex system is spatially asymmetric, thus
breaking the symmetry of the full Hamiltonian \cite{RE}). Other
representations of the same symmetry may be realized by the system's excited
states. Thus, the GS wave function trapped in the one-dimensional DWP is
symmetric (even) with respect to the double-well structure, while the first
excited state is antisymmetric (odd).

Unlike the linear quantum-mechanical Schr\"{o}dinger equation for a single
particle, atomic Bose-Einstein condensates (BECs) are modeled by the
Gross-Pitaevskii equation (GPE) for the single-atom wave function $\psi
\left( x,t\right) $, with a trapping potential, $U(x)$, and the cubic term
which accounts for collisions between atoms, in the framework of the
mean-field approximation \cite{BEC}:

\begin{equation}
i\frac{\partial \psi }{\partial t}=-\frac{1}{2}\frac{\partial ^{2}\psi }{%
\partial x^{2}}-g\left\vert \psi \right\vert ^{2}\psi +U(x)\psi .
\label{psi}
\end{equation}%
This equation is written in the scaled form, which can be derived from the
three-dimensional GPE for the cigar-shaped configuration, with strong
confinement applied in the transverse plane \cite{Luca}. The repulsive or
attractive interactions between atoms correspond, respectively, to the
self-defocusing ($g<0$) or self-focusing ($g>0$) sign of the cubic term in
Eq. (\ref{psi}).\ Similarly, the nonlinear Schr\"{o}dinger equation (NLSE)
with the cubic term governs the paraxial propagation of electromagnetic
waves in optical media with the Kerr nonlinearity \cite{NLS}. In the latter
case, Eq. (\ref{psi}) applies to the light transmission in the spatial
domain, with $t$ replaced by the propagation distance, $z$, and $-U(x)$
representing a transverse modulation profile of the local refractive index,
which imposes a guiding structure in the $\left( x,z\right) $ plane.

Many models of nonlinear optics and BEC are based on GPE/NLSE in the form of
Eq. (\ref{psi}) with potential $U(x)$ representing symmetric DWPs. A
fundamental difference from the linear systems is that the GS in the
self-focusing models follows the symmetry of the underlying potential
structure only if the nonlinearity remains relatively weak. A generic
effect, which occurs with the increase of the strength of the nonlinearity
as a result of its interplay with the DWP, is \textit{spontaneous symmetry
breaking} (SSB), which destabilizes the symmetric GS and replaces it by an
asymmetric one \cite{book}. The switch from the symmetric GS to its
asymmetric counterpart occurs via the corresponding \textit{bifurcation}
(phase transition) at a critical value of the nonlinearity strength \cite%
{Iooss,Kuzn}.

Originally, the SSB was predicted in simple models based on systems of
linearly coupled equations with intrinsic nonlinearity \cite{old}. In
particular, Eq. (\ref{psi}) with potential $U(x)$ in the form of two
symmetric deep wells can be reduced to a system of two coupled ordinary
differential equations for amplitudes\ $u_{1}(t)$ and $u_{2}(t)$ in the
framework of the tight-binding approximation \cite{tight}, which replaces $%
\psi (x,t)$ by a superposition of two stationary wave functions, $\phi $,
corresponding to the states trapped separately in the two deep potential
wells, centered at $x=\pm a$:
\begin{equation}
\psi \left( x,t\right) =u_{1}(t)\phi \left( x-a\right) +u_{2}(t)\phi \left(
x+a\right) .  \label{u1u2}
\end{equation}

In terms of the BEC, the SSB gives rise to the GS with the atomic density in
one well of the trapping DWP larger than in the other. The SSB also breaks
another principle of quantum mechanics, according to which the GS cannot be
degenerate, as there emerge a degenerate pair of mirror-image asymmetric GS
wave functions, with the larger density self-trapped in either of the two
potential wells. In optics, the SSB means that larger light power
spontaneously self-traps in either core of the nonlinear dual-core waveguide.

Thus, the SSB effect is common to diverse systems which combine the wave
transmission, self-focusing, and trapping potentials of the DWP type. In
photonics, the SSB was reported in several experimental works. In
particular, the symmetry breaking for a pair of laser beams coupled into a
transverse DWP created in a self-focusing photorefractive medium was
demonstrated in Ref. \cite{photo}. Another experimental result is a
spontaneously established asymmetric regime of the operation of a symmetric
pair of coupled lasers \cite{lasers}. More recently, SSB was demonstrated in
a pair of nanolaser cavities embedded into a photonic crystal \cite{France}.
Observation of spontaneous breaking of the chiral symmetry in metamaterials
was reported in Ref. \cite{Kivshar}.

The analysis of the SSB in the model based on Eq. (\ref{psi}) was initiated
in Refs. \cite{Milburn} and \cite{Smerzi}. Most often, the BEC\ nonlinearity
(on the contrary to the self-focusing Kerr effect in optics) is
self-repulsive, which corresponds to $g<0$ in Eq. (\ref{psi}). In this case,
the symmetric (even) GS is not subject to destabilization, but the first
antisymmetric (odd) excited state, with $\psi (-x)=-\psi (x)$, suffers
destabilization and spontaneous breakup of its antisymmetry when the
strength of the repulsive nonlinearity attains a critical level \cite{book}.
This manifestation of the SSB phenomenology was demonstrated experimentally
in Ref. \cite{Markus}, using the condensate of $^{87}$Rb atoms with
repulsive interactions between them, loaded into a DWP trapping
configuration.

The above discussion addressed static symmetric and asymmetric modes in the
nonlinear systems including the DWP structure. Dynamical regimes, in the
form of oscillations of the wave function between two wells of the DWP,
i.e., roughly speaking, between the two mirror-image asymmetric states
existing above the SSB point, were studied too. Following the analogy with
Josephson oscillations in tunnel-coupled superconductors \cite%
{superconductor,Ustinov}, the possibility of the matter-wave oscillations in
\textit{bosonic Josephson junctions} was predicted \cite{junction} and
experimentally realized in the trapped BEC \cite{Markus}.

Additional dynamical regimes were studied in Ref. \cite{Greeks}, in the
framework of a model which combines the DWP and ``nonlinearity management",
i.e., time-periodic modulation of coefficient $g$ in Eq. (\ref{psi}). It was
demonstrated that the symmetry-breaking dynamics may be strongly altered by
the application of the management: the SSB can be suppressed in cases when
it occurs, and induced in cases when it does not take place in the absence
of the management.

\begin{figure}[tbp]
\includegraphics[width=3.2in]{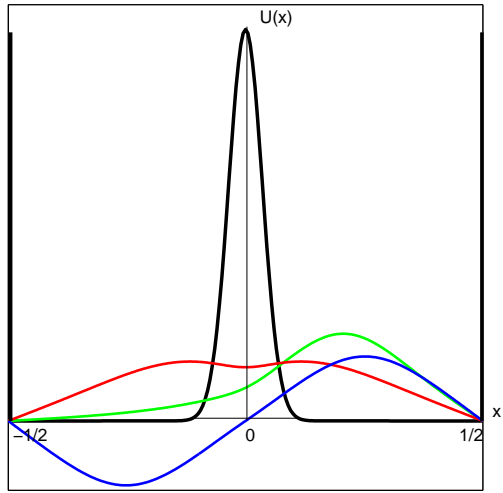}
\caption{{}The setting under the consideration. An infinitely deep potential
box of width $1$ is split in the middle by a narrow tall barrier,
approximated by $U(x)=\protect\varepsilon \protect\delta (x)$, see Eq. (%
\protect\ref{psi2}). Wave functions of symmetric (even), antisymmetric
(odd), and asymmetric stationary states are schematically shown by red,
blue, and green curves, respectively.}
\label{fig1}
\end{figure}

The objective of the present work is to explore, by means of analytical and
numerical methods, the SSB\ in what may be considered as the most
fundamental version of the systems represented by Eq. (\ref{psi}) with the
self-attractive nonlinearity ($g>0$), namely, an infinitely deep potential
box (whose width is scaled to be $1$), split in two wells by a narrow
barrier, as schematically shown in Fig. \ref{fig1}. In this case, Eq. (\ref%
{psi}) with the barrier represented by the ideal $\delta $-function takes
the form of the following equation subject to zero boundary conditions:%
\begin{gather}
i\frac{\partial \psi }{\partial t}=-\frac{1}{2}\frac{\partial ^{2}\psi }{%
\partial x^{2}}-g\left\vert \psi \right\vert ^{2}\psi +\varepsilon \delta
(x)\psi ,  \label{psi2} \\
\psi \left( x=\pm \frac{1}{2}\right) =0,  \label{b.c.}
\end{gather}%
where $\varepsilon >0$ is the strength of the splitter, and $g=1$ may be
fixed, unless $g=0$ in the linear version of Eq. (\ref{psi2}). The
Hamiltonian (energy) corresponding to Eq. (\ref{psi2}) and (\ref{b.c.}) is%
\begin{equation}
H=\frac{1}{2}\int_{-1/2}^{+1/2}\left( \left\vert \frac{\partial \psi }{%
\partial x}\right\vert ^{2}-\left\vert \psi \right\vert ^{4}\right)
dx+\varepsilon \left\vert \psi (x=0\right\vert ^{2}.  \label{H}
\end{equation}%
In a sense, this model is opposite to the one introduced in recent work \cite%
{Krzy}, in which the SSB was studied in a DWP with an \textquotedblleft
elevated floor" (rather than the infinitely deep one), i.e., a DWP structure
embedded into a broad potential barrier.

It is relevant to mention that, although the one-dimensional NLSE is
integrable in the free space, the internal potential and boundary conditions
in Eqs. (\ref{psi2}) and (\ref{b.c.}) destroy the integrability, therefore
the evolution of unstable states in this model is not expected to be
periodic or quasi-periodic in time, see Figs. \ref{fig12} and \ref{fig15}(c)
below as examples. The model gives rise to nearly integrable dynamics only
in the case when the solution may be approximated by a set of narrow
solitons (this case is considered in Section II.C). It may be expected that
the dynamics reduced to the variational approximation (VA), based on ansatz (%
\ref{ans}) with three degrees of freedom, may be close to quasi-periodic, as
the presence of two dynamical invariants, \textit{viz}., the norm, given by
Eq. (\ref{abc}), and the respective Hamiltonian, makes the approximate
dynamical system nearly integrable, with quasi-periodic trajectories carried
by KAM tori \cite{KAM}. Actually, our VA-based analysis is focused below on
its stationary version, which is most essential in terms of the present
work. On the other hand, it should be stressed that the nonintegrable
dynamics remains reversible \cite{KAM}, as the model does not include any
dissipation.

In spite of the fact that Eqs. (\ref{psi2})-(\ref{H}) present, as a matter
of fact, the simplest version of model (\ref{psi}), it was introduced only
recently in Refs. \cite{NatPhot} and \cite{ClercTlidi} (where some
analytical approximations were introduced, but systematic analysis was not
performed). The model can be implemented in BEC, using strong repelling
optical sheets to emulate both the outer potential walls and the inner
barrier, the widths of the latter and of the whole potential box being,
respectively $\sim 1$ $\mathrm{\mu }$m and $\sim 100$ $\mathrm{\mu }$m in
physical units. In optics, the box structure may be realized as a step-index
one made in silica \cite{SnyderLove}. Assuming the width of the waveguide $%
\sim 20$ $\mathrm{\mu }$m (to have it narrow enough for potential
applications), characteristic values of interest, $\varepsilon \sim 10$ (see
below), correspond to the splitting layer of thickness $\sim 2$ $\mathrm{\mu
}$m.

The rest of the paper is organized as follows. An analytical approach to the
model is recapitulated in Section II, following its (incomplete)
introduction in Ref. \cite{ClercTlidi}, which is necessary for comparison
with numerical results. The analytical part includes the VA for stationary
symmetric and asymmetric modes, as well as approximations for weakly and
strongly split potential boxes, i.e., small and large $\varepsilon $ , in
terms of Eq. (\ref{psi2}). Results of systematic numerical studies and the
comparison with analytical predictions are reported in Section III, which
demonstrates that the VA is quite accurate for the model with a relatively
tall splitting barrier ($\varepsilon $ not too small), while the VA yields a
wrong bifurcation diagram for small $\varepsilon $, due to inadequacy of the
underlying ansatz (when, however, the other analytical approximation, based
on the soliton ansatz, applies correctly). The numerical results are
reported for symmetric, asymmetric, and antisymmetric bound states (the
latter one representing the first excited state in the system), including
simulations of dynamical regimes initiated by the SSB instability of
symmetric and antisymmetric ones. The paper is concluded by Section IV.

\section{The analytical approach}

\subsection{Stationary modes}

Stationary states produced by Eqs. (\ref{psi2}) and (\ref{b.c.}) with real
energy eigenvalue $\mu $ (in terms of optics, $-\mu $ is the propagation
constant)\ are looked for as
\begin{equation}
\psi (x,t)=e^{-i\mu t}\phi (x),  \label{stat}
\end{equation}%
with real wave function $\phi (x)$ obeying a stationary equation with the
respective b.c.:
\begin{equation}
\mu \phi =-\frac{1}{2}\frac{d^{2}\phi }{dx^{2}}-g\phi ^{3}+\varepsilon
\delta (x)\phi ,~\phi \left( x=\pm \frac{1}{2}\right) =0.  \label{NLSE}
\end{equation}%
The presence of the delta-functional barrier at $x=0$ implies that $\phi (x)$
is continuous at this point, while its derivative obeys a jump condition:%
\begin{equation}
\frac{d\phi }{dx}|_{x=+0}-\frac{d\phi }{dx}|_{x=-0}=2\varepsilon \phi (x=0).
\label{jump}
\end{equation}

As said above, $g\equiv 1$ is fixed throughout the paper, unless $g=0$ is
set in the linear case, hence the strength of the nonlinearity is determined
by the norm of the wave function,%
\begin{equation}
N=\left( \int_{-1/2}^{0}+\int_{0}^{+1/2}\right) \phi ^{2}(x)dx\equiv
N_{-}+N_{+}.  \label{N}
\end{equation}%
The asymmetry of states above the SSB point is characterized by the relative
difference of the norms in the right and left sections of the split
potential box:%
\begin{equation}
\Theta \equiv \left( N_{+}-N_{-}\right) /\left( N_{+}+N_{-}\right) .
\label{theta}
\end{equation}

Before proceeding to the consideration of the nonlinear model, it is
relevant to dwell on its linear counterpart, with $g=0$ in Eq. (\ref{NLSE}).
Spatially symmetric (even) solutions of the linear equation are looked for as%
\begin{equation}
\phi _{\mathrm{even}}^{(\mathrm{lin})}(x)=A\sin \left( \sqrt{2\mu }\left(
\frac{1}{2}-|x|\right) \right) ,  \label{linear}
\end{equation}%
where $A$ is an arbitrary amplitude, and eigenvalue $\mu $ is determined by
a relation following from Eq. (\ref{jump}):%
\begin{equation}
\tan \left( \sqrt{\mu /2}\right) =-\sqrt{2\mu }/\varepsilon .  \label{eigen}
\end{equation}%
It is easy to see that, with the increase of $\varepsilon $ from $0$ to $%
\infty $, the lowest eigenvalue, $\mu _{0}$, which corresponds to the GS of
the linear model, monotonously grows from $\mu _{0}\left( \varepsilon
=0\right) =\pi ^{2}/2$ to
\begin{equation}
\mu _{0}\left( \varepsilon =\infty \right) =2\pi ^{2}.  \label{muGS}
\end{equation}%
Similarly, the eigenvalue of the first excited symmetric state, $\mu
_{2}\left( \varepsilon \right) $, monotonously grows from $\mu _{2}\left(
\varepsilon =0\right) =9\pi ^{2}/2$ to $\mu _{2}\left( \varepsilon =\infty
\right) =8\pi ^{2}$. Eigenvalue $\mu _{1}=2\pi ^{2}$, which corresponds to
the lowest excited state, i.e., the first antisymmetric (odd) eigenfunction,
\begin{equation}
\phi _{\mathrm{odd}}^{(\mathrm{lin})}(x)=A\sin \left( \sqrt{2\mu _{1}}%
x\right)   \label{odd}
\end{equation}%
[$\mu _{1}$ does not depend on $\varepsilon $, as wave function (\ref{odd})
vanishes at $x=0$], is located between $\mu _{0}$ and $\mu _{2}$. Naturally,
$\mu _{1}$ coincides with limit value (\ref{muGS}) of $\mu _{0}$, as the GS
eigenfunction also vanishes at $x=0$, in the limit of $\varepsilon =\infty $.

\subsection{The deeply-split double-well-potential (large $\protect%
\varepsilon $)}

The first objective of the analysis of the nonlinear model based on Eq. (\ref%
{NLSE}) is to predict the critical norm at the SSB point. In the case of
weakly coupled (deeply split) potential wells, which corresponds to large $%
\varepsilon $ (a very tall splitting barrier), weak nonlinearity, which
corresponds to a small norm and amplitude of the wave function, is
sufficient to induce the SSB in the competition with the weak linear
coupling. The small amplitude implies that relevant solutions to Eq. (\ref%
{NLSE}) are close to eigenmodes (\ref{linear}) of the linear equation, hence
an approximate solution may be looked for as%
\begin{equation}
\phi (x)=A_{\pm }\sin \left( k_{\pm }\left( \frac{1}{2}-|x|\right) \right) ,
\label{lin}
\end{equation}%
where signs $\pm $ pertain to $x<0$ and $x>0$, respectively. The
substitution of this \textit{ansatz} into the condition of the continuity of
the wave function at $x=0$, and relation (\ref{jump}) for its first
derivative, leads to the following relations between amplitudes $A_{\pm }$
and the wavenumbers:%
\begin{gather}
A_{+}\sin \left( \frac{1}{2}k_{+}\right) =A_{-}\sin \left( \frac{1}{2}%
k_{-}\right) ,  \label{bc1} \\
A_{-}k_{-}\cos \left( \frac{1}{2}k_{-}\right) -A_{+}k_{+}\cos \left( \frac{1%
}{2}k_{+}\right) =4\varepsilon A_{\pm }\sin \left( \frac{1}{2}k_{\pm
}\right) .  \label{bc2}
\end{gather}%
In the same small-amplitude approximation, the third harmonic contained in
the cubic term in Eq. (\ref{NLSE}) may be neglected, hence this term is
approximated as
\begin{equation}
\left[ A_{\pm }\sin \left( k_{\pm }\left( \frac{1}{2}-|x|\right) \right) %
\right] ^{3}\approx \frac{3}{4}A_{\pm }^{3}\sin \left( k_{\pm }\left( \frac{1%
}{2}-|x|\right) \right) .  \label{AA}
\end{equation}%
Further, Eq. (\ref{AA}) implies that the cubic term amounts to an effective
shift of the energy eigenvalue in Eq. (\ref{NLSE}), which determines the
wavenumbers in Eq. (\ref{lin}):%
\begin{equation}
k_{\pm }=\sqrt{2\left( \mu +\frac{3}{4}gA_{\pm }^{2}\right) }.  \label{k}
\end{equation}

In the limit of $\varepsilon \rightarrow \infty $, wave functions (\ref{lin}%
) must vanish at $x=0$, hence the respective GS corresponds to $k_{\pm
}=2\pi $. At large but finite $\varepsilon $, the energy corresponding to
the GS should be close to the limit value given by Eq. (\ref{muGS}),%
\begin{equation}
\mu =2\pi ^{2}-\delta \mu ,~\delta \mu \ll 2\pi ^{2}.  \label{delta-mu}
\end{equation}%
The substitution of expression (\ref{delta-mu}) into Eq. (\ref{k}),
expanding it for small $\delta \mu $ and $A_{\pm }^{2}$, inserting the
result into Eqs. (\ref{bc1}) and (\ref{bc2}), and taking the limit of $%
A_{+}-A_{-}\rightarrow 0$, which corresponds to the SSB bifurcation point,
leads to the following analytical prediction for the bifurcation-point
parameters:%
\begin{equation}
N_{\mathrm{bif}}=\frac{8}{3}\pi ^{2}\varepsilon ^{-1},~\left( A_{\pm
}^{2}\right) _{\mathrm{bif}}=2N_{\mathrm{cr}},~\delta \mu =12\pi
^{2}\varepsilon ^{-1}.  \label{cr}
\end{equation}%
Thus, as expected, the value of the norm at the SSB point decays ($\sim
\varepsilon ^{-1}$) with the increase of $\varepsilon $. This approximate
analytical result is compared with its numerical counterpart in Fig. \ref%
{fig2}(a) (further numerical results are presented in the next section).
\begin{figure}[tbp]
\subfigure[]{\includegraphics[width=3.2in]{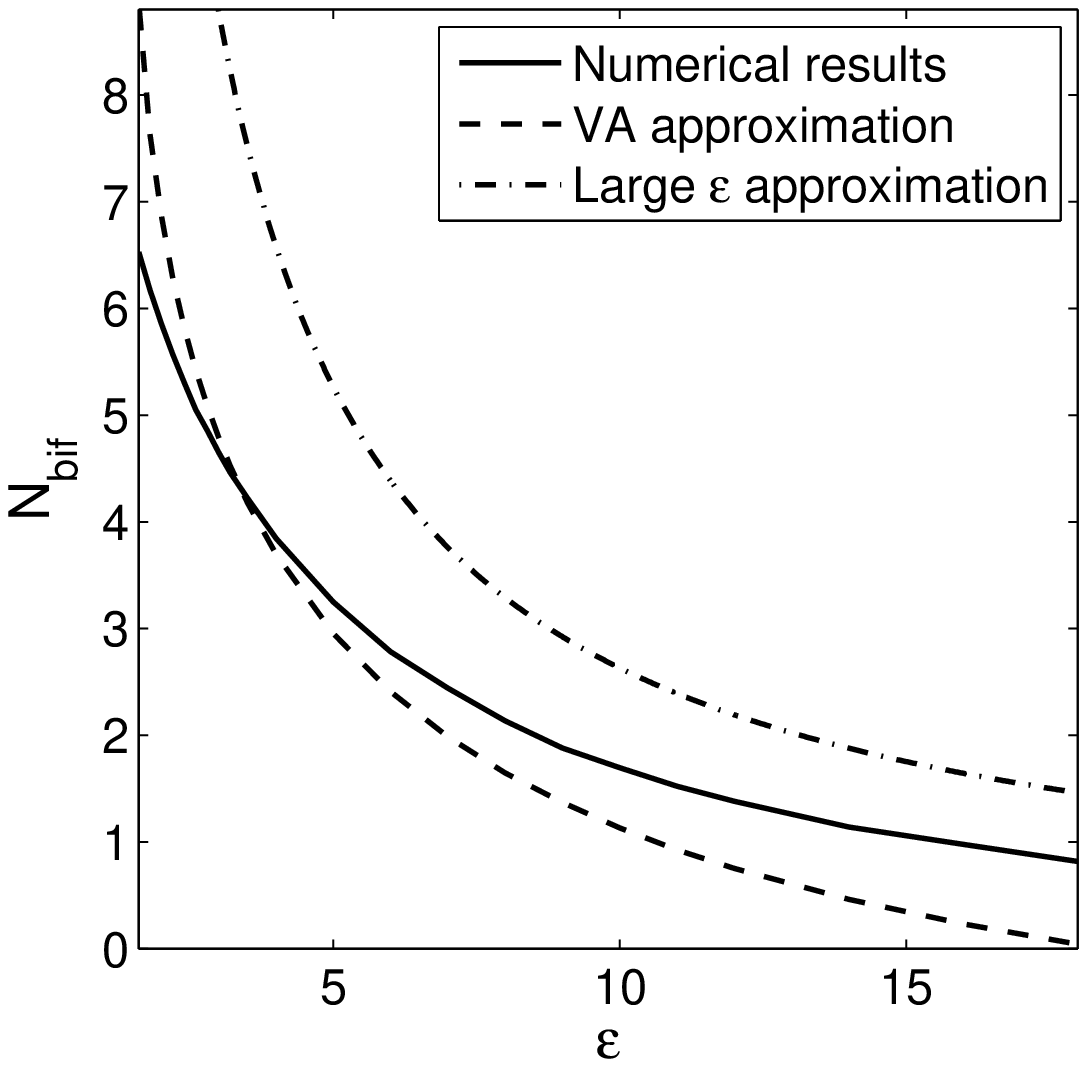}} \subfigure[]{%
\includegraphics[width=3.2in]{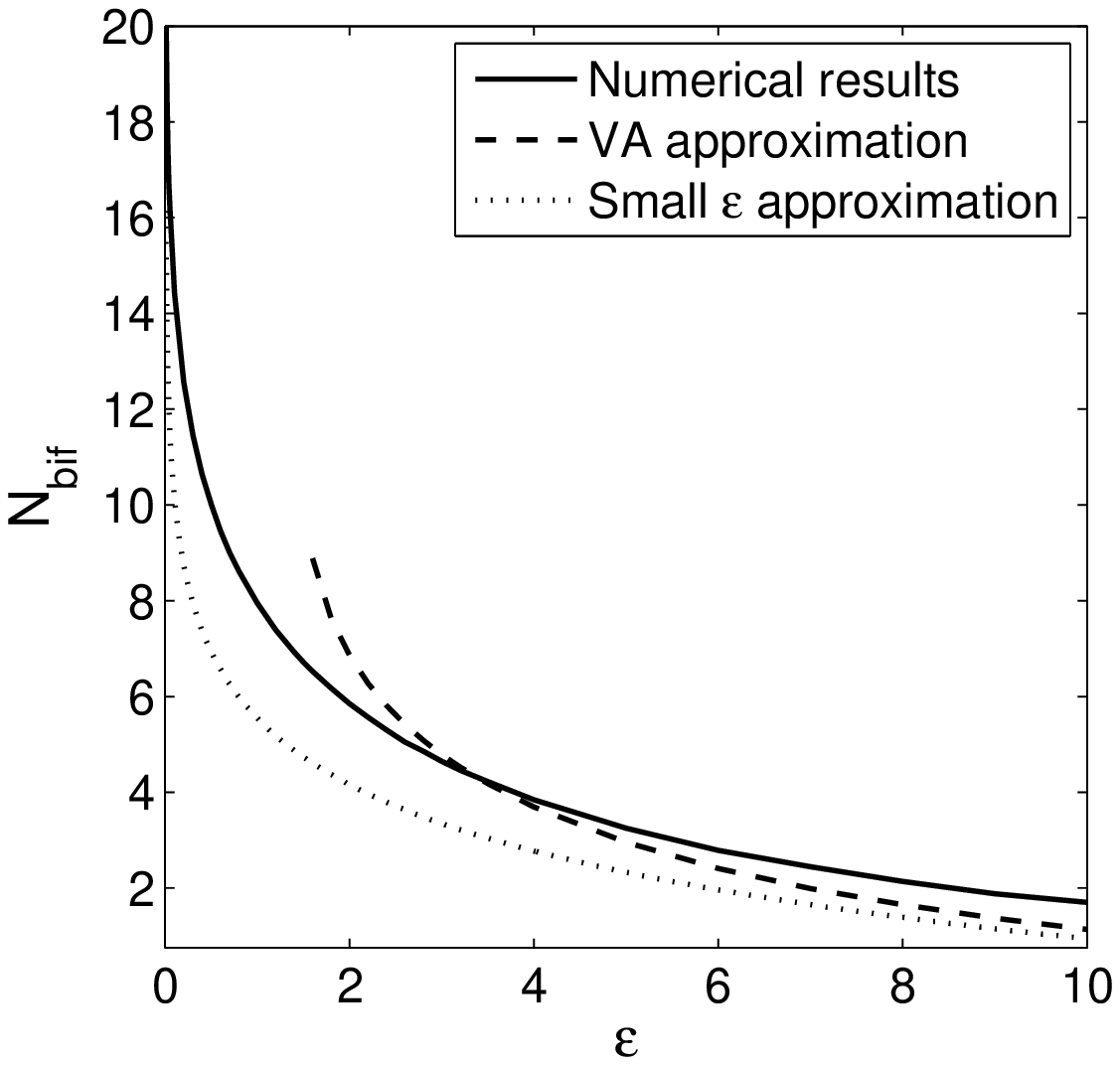}}
\caption{{}The norm at the symmetry-breaking bifurcation point versus the
strength of the splitting barrier, $\protect\varepsilon $. Along with the
numerically found and VA-predicted dependences, the analytical
approximations based on Eqs. (\protect\ref{cr}) and (\protect\ref{Ncr}),
which are relevant, respectively, for large and small $\protect\varepsilon $%
, are shown in panels (a) and (b) [note that in (a) the range starts from $%
\protect\varepsilon \approx 2.7$].}
\label{fig2}
\end{figure}

\subsection{The weakly-split double-well-potential (small $\protect%
\varepsilon $)}

Small $\varepsilon $ implies strong coupling of the two potential wells with
a shallow split between them, therefore strong nonlinearity, i.e., large $N$%
, is required to cause the SSB under the competition with the strong
coupling. In turn, large $N$ implies that the wave field tends to self-trap
into a narrow NLSE\ soliton \cite{NLS},%
\begin{equation}
\phi _{\mathrm{sol}}\left( x-\xi \right) =\frac{1}{2}N\mathrm{sech}\left(
\frac{1}{2}N\left( x-\xi \right) \right) ,  \label{sol}
\end{equation}%
where $\xi $ is the coordinate of the soliton's center, the respective
energy eigenvalue being%
\begin{equation}
\mu _{\mathrm{sol}}=-N^{2}/8.  \label{mu-sol}
\end{equation}%
This approximation is valid provided that the soliton's width is much
smaller than the size of the potential box, which means $N\gg 1$.

According to the b.c. in Eq. (\ref{NLSE}), the soliton interacts with two
\textit{ghost solitons}, i.e., its mirror images (with opposite signs), with
respect to the edges of the box:%
\begin{equation}
\phi _{\mathrm{ghost}}(x)=-\frac{N}{2}\left[ \mathrm{sech}\left( \frac{1}{2}%
N\left( x-1+\xi \right) \right) +\mathrm{sech}\left( \frac{1}{2}N\left(
x+1+\xi \right) \right) \right] .  \label{ghost}
\end{equation}%
The sum of the well-known potentials \cite{sol-sol,RMP} of the interaction
of the given soliton with the two ``ghosts" gives rise to an
effective potential of repulsion of the real soliton from edges of the
confining box:%
\begin{equation}
U_{\mathrm{box}}(\xi )=N^{3}\exp \left( -N/2\right) \cosh \left( N\xi
\right) .  \label{box}
\end{equation}

On the other hand, the same soliton is repelled by the splitting barrier,
with the corresponding potential \cite{RMP}%
\begin{equation}
U_{\mathrm{barrier}}(\xi )=\varepsilon \phi _{\mathrm{sol}}^{2}\left( \xi
=0\right) =\frac{\varepsilon }{4}N^{2}\mathrm{sech}^{2}\left( \frac{1}{2}%
N\xi \right) ,  \label{barrier}
\end{equation}%
if the deformation of the soliton's shape by the weak barrier is neglected.
A straightforward analysis of the total effective potential, $U(\xi )=U_{%
\mathrm{box}}(\xi )+U_{\mathrm{barrier}}(\xi )$, demonstrates that the
position of the soliton placed at $\xi =0$, which represents the symmetric
mode in the present case, is stable, i.e., it corresponds to a \emph{minimum}
of the net potential, at $8N\exp \left( -N/2\right) >\varepsilon ,$ or, in
other words, at%
\begin{equation}
N<N_{\mathrm{bif}}\approx 2\ln \left( 16/\varepsilon \right) .  \label{Ncr}
\end{equation}%
The SSB bifurcation takes place, with the increase of $N$, at $N=N_{\mathrm{%
bif}}$, when the local potential minimum at $\xi =0$ switches into a
maximum. At $0<\left( N-N_{\mathrm{bif}}\right) /N_{\mathrm{bif}}\ll 1$, the
center of the soliton spontaneously shifts to either of two asymmetric
positions, which correspond to a pair of new potential minima emerging at $%
\xi =\pm \sqrt{\left( N-N_{\mathrm{bif}}\right) }/N_{\mathrm{cr}}$.

The comparison of the approximate analytical result given by Eq. (\ref{Ncr})
with its numerically obtained counterpart is displayed in Fig. \ref{fig2}%
(b). At moderate values of $\varepsilon $ the discrepancy is relatively
large because the validity condition for the present approximation is that $%
\varepsilon $ must be so small that $\ln \left( 16/\varepsilon \right) $ may
be treated as a large parameter.

\subsection{The variational approximation (VA)}

In the generic case, when the strength of the splitting barrier, $%
\varepsilon $, is not assumed to be specifically large or small, an
analytical consideration may be based on the VA \cite{Progress}, which is
suggested by the fact that stationary equation (\ref{NLSE}) may be derived
from the minimization of the corresponding Lagrangian,%
\begin{equation}
L=\int_{-1/2}^{+1/2}\left[ \frac{1}{2}\left( \frac{d\phi }{dx}\right)
^{2}-\mu \phi ^{2}-\frac{g}{2}\phi ^{4}\right] dx+\varepsilon \phi
^{2}(x=0)\equiv H-\mu N,  \label{L}
\end{equation}%
where $H$ and $N$ are the Hamiltonian (\ref{H}) and norm (\ref{N}) defined
above. The following \textit{ansatz} for the GS wave function, satisfying
b.c. in Eq. (\ref{NLSE}), is the simplest one which is capable to capture
the SSB:

\begin{equation}
\phi (x)=a\cos (\pi x)+b\sin (2\pi x)+c\cos (3\pi x),  \label{ans}
\end{equation}%
cf. Eq. (\ref{u1u2}), where real amplitudes $a,$ $c,$ and $b$ must be
predicted by the VA. The SSB is accounted for by the odd term $\sim b$ in
the ansatz, which breaks the symmetry of the even expression. Accordingly,
the onset of the SSB is signaled by the emergence of a solution with
infinitesimal $b$, branching off from from the symmetric one with $b=0$.

The integral norm (\ref{N}) of ansatz (\ref{ans}) is
\begin{equation}
N=\left( 1/2\right) \left( a^{2}+c^{2}+b^{2}\right) ,  \label{abc}
\end{equation}
while its asymmetry at $b\neq 0$ is quantified by parameter (\ref{theta}),%
\begin{equation}
\Theta =\frac{16}{15\pi }\frac{b\left( 5a-3c\right) }{a^{2}+c^{2}+b^{2}}.
\label{Theta}
\end{equation}%
A straightforward consideration confirms that expression (\ref{Theta}) is
always subject to constraint $\left\vert \Theta \right\vert <1$, as it must
be. The Sturm theorem, according to which the spatially symmetric GS cannot
have nodes \cite{LL}, i.e., $\phi (x)\neq 0$ at $|x|<1/2$ (it remains valid
in the nonlinear system), if applied to ansatz (\ref{ans}) with $b=0$,
amounts to the following constraint:%
\begin{equation}
-1<c/a<1/3.  \label{or}
\end{equation}

The substitution of ansatz (\ref{ans}) into Lagrangian (\ref{L}) yields

\begin{gather}
L_{\mathrm{VA}}=\allowbreak \left( \frac{1}{4}\pi ^{2}-\frac{1}{2}\mu
+\varepsilon \right) a^{2}+\left( \pi ^{2}-\frac{1}{2}\mu \right)
b^{2}+\left( \frac{9}{4}\pi ^{2}-\frac{1}{2}\mu +\varepsilon \right)
c^{2}\allowbreak +2\varepsilon ac  \notag \\
-\frac{1}{4}\left( \frac{3}{4}a^{4}+a^{3}c+3a^{2}b^{2}+3a^{2}c^{2}-3ab^{2}c+%
\frac{3}{4}\allowbreak b^{4}+3b^{2}c^{2}+\frac{3}{4}c^{4}\right) ,
\label{LL}
\end{gather}%
which gives rise to the variational equations for real amplitudes $b$ and $%
a,c$:
\begin{equation}
\partial L/\partial (b^{2})=0,  \label{d/db}
\end{equation}%
\begin{equation}
\partial L/\partial a=\partial L/\partial c=0.  \label{dd/dadc}
\end{equation}%
To identify the bifurcation point at which the SSB sets in, one may set $b=0$
in Eqs. (\ref{d/db}) and (\ref{dd/dadc}) [\emph{after} performing the
differentiation with respect to $b^{2}$ in Eq. (\ref{d/db})]. This procedure
leads to a system of equations for the values of $a$, $b$, and $\mu $ at the
bifurcation point:
\begin{equation}
2\pi ^{2}-\mu =\frac{3}{2}\left( a^{2}-ac+c^{2}\right) ,  \label{bsimple}
\end{equation}%
\begin{gather}
\left( \frac{1}{2}\pi ^{2}-\mu +2\varepsilon \right) a+2\varepsilon c-\frac{1%
}{4}\left( 3a^{3}+3a^{2}c+6ac^{2}\right) =0,  \label{asimple} \\
\left( \frac{9}{2}\pi ^{2}-\mu +2\varepsilon \right) c\allowbreak
+2\varepsilon a-\frac{1}{4}\left( a^{3}+6a^{2}c+3c^{3}\right) =0.
\label{csimple}
\end{gather}%
It is easy to check that Eqs. (\ref{bsimple})-(\ref{csimple}) have no
relevant solutions at $\varepsilon =0$, in agreement with the obvious fact
that the SSB does not occur when the central barrier is absent, i.e., the
potential box is not split into two wells.

The VA-produced prediction for the bifurcation point, in the form of $N_{%
\mathrm{bif}}(\varepsilon )$, obtained from a numerical solution of Eqs. (%
\ref{bsimple})-(\ref{csimple}), is compared to its numerically found
counterpart in Figs. \ref{fig2}(a,b). It is seen that the VA provides a
reasonable accuracy at moderate values of $\varepsilon $. At large $%
\varepsilon $, the discrepancy is explained by the fact that ansatz (\ref%
{ans}) does not take into regard the derivative's jump given by Eq. (\ref%
{jump}). On the other hand, at very small $\varepsilon $, the relevant
ansatz is also different, as it amounts to soliton (\ref{sol}).

\begin{figure}[tbp]
\includegraphics[width=3.2in]{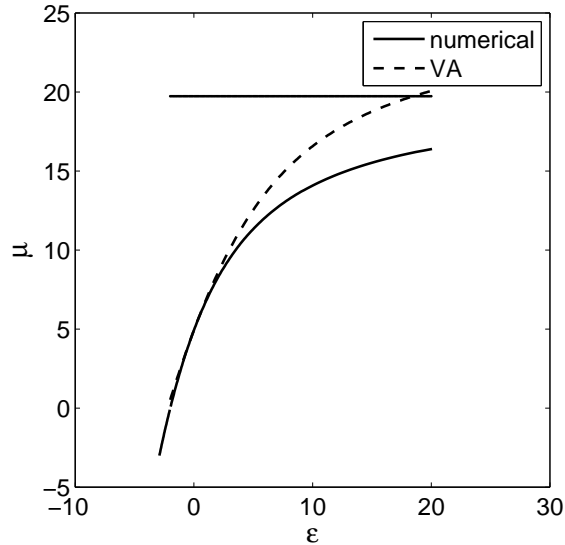}
\caption{{}The eigenvalue for the energy of the GS\ (ground state) in the
linear version of Eq. (\protect\ref{NLSE}) ($g=0$) versus the strength of
the splitting barrier, $\protect\varepsilon $, as obtained from the
numerical solution of Eq. (\protect\ref{eigen}), and as predicted by the VA
in the form of Eq. (\protect\ref{GS}). The horizontal line shows the
constraint imposed by Eq. (\protect\ref{muGS}).}
\label{fig3}
\end{figure}

Equations (\ref{asimple}) and (\ref{csimple}) in which the cubic terms are
dropped [and Eq. (\ref{bsimple}) is dropped too] correspond to the
linearized version of Eq. (\ref{NLSE}), with $g=0$. This simplest version of
the VA predicts the above-mentioned eigenvalue, $\mu _{0}(\varepsilon )$,
which corresponds to the GS of the linear system, as a value at which the
determinant of the linearized version of Eqs. (\ref{asimple}) and (\ref%
{csimple}) for $\ a$ and $c$ vanishes:
\begin{equation}
\left( \mu _{0}\right) _{\mathrm{VA}}=\frac{5}{2}\pi ^{2}+2\varepsilon -2%
\sqrt{\pi ^{4}+\varepsilon ^{2}}  \label{GS}
\end{equation}%
[recall that Eq. (\ref{eigen}) for $\mu (\varepsilon )$ cannot be solved
exactly, except for concluding that $\mu _{0}$ is a monotonously growing
function of $\varepsilon $ confined to interval $\pi ^{2}/2\leq \mu
_{0}<2\pi ^{2}$, see Eqs. (\ref{muGS}) and (\ref{delta-mu})]. In particular,
the approximate GS energy, produced by Eq. (\ref{GS}), satisfies the
constraint $\mu _{0}<2\pi ^{2}$ at $\varepsilon <\left( 15/8\right) \pi
^{2}\approx 18.5$. This approximate result is compared to its numerical
counterpart in Fig. \ref{fig3}. As said above, the discrepancy at large $%
\varepsilon $ is explained by the fact that ansatz (\ref{ans}) does not take
into regard the jump of the derivative at $x=0$. The predictions of the VA
are further compared with numerical findings in the next section.


\section{Numerical results}

Numerical solutions of Eqs. (\ref{psi2}) and (\ref{NLSE}) were obtained
replacing the ideal $\delta $-function by its regularized version,%
\begin{equation}
\tilde{\delta}(x)=\frac{1}{\sqrt{\pi }\xi }\exp \left( -\frac{x^{2}}{\xi ^{2}%
}\right) ,  \label{delta}
\end{equation}%
with width $\xi \ll 1$ (the results are presented below for $\xi =0.05$; at
still smaller values of the regularization width, such as $\xi =0.01$, the
findings are essentially the same). Solutions of the stationary equation (%
\ref{NLSE}) were produced by means of the Newton's method, with mesh size $%
\Delta x=1/1025$. Direct simulations of the evolution governed by Eq. (\ref%
{psi2}) were performed by dint of the standard split-step algorithm, with
time step $\Delta t=0.001$.

The stability of the stationary modes was explored through a numerical
solution of the GPE linearized for small perturbations around the stationary
mode. For this purpose, the perturbed version of stationary solutions (\ref%
{stat}) is taken in the usual form:%
\begin{equation}
\psi \left( x,t\right) =e^{-i\mu t}\left\{ \phi (x)+\eta \left[ e^{-i\lambda
t}u(x)+e^{i{\lambda }^{\ast }t}v^{\ast }(x)\right] \right\} ,  \label{pert}
\end{equation}%
where $\eta $ is an infinitesimal amplitude of the perturbations, $u(x)$ and
$v(x)$ represent an eigenmode, and $\lambda $ is the corresponding
(generally, complex) perturbation eigenfrequency, the stability condition
being $\mathrm{Im}\{\lambda \}=0$ for all $\lambda $. The substitution of
expression (\ref{pert}) in Eqs. (\ref{psi2}) and (\ref{b.c.}) and the
subsequent linearization gives rise to the eigenvalue problem for $\lambda $%
, based on the following equations:%
\begin{gather}
\left( \mu +\lambda \right) u+\frac{1}{2}u^{\prime \prime }+g\phi
^{2}(x)\left( 2u+v\right) =\varepsilon \delta (x)u,  \notag \\
\left( \mu -\lambda \right) v+\frac{1}{2}v^{\prime \prime }+g\phi
^{2}(x)\left( 2v+u\right) =\varepsilon \delta (x)v,  \notag \\
u\left( x=\pm \frac{1}{2}\right) =v\left( x=\pm \frac{1}{2}\right) =0,
\label{linearized}
\end{gather}%
with the prime standing for $d/dx$. Equations (\ref{linearized}) were solved
numerically [with $g=1$ and $\delta (x)$ replaced by $\tilde{\delta}(x)$, as
per Eq. (\ref{delta})] by means of the finite-difference method. Predictions
for (in)stability of stationary modes, produced by the linearized GPE, were
verified by means of direct simulations of their perturbed evolution in the
framework of full equation (\ref{psi2}).

\subsection{Symmetric and asymmetric modes and the SSB at moderate values of
$\protect\varepsilon $}

\begin{figure}[tbp]
\subfigure[]{\includegraphics[width=3.2in]{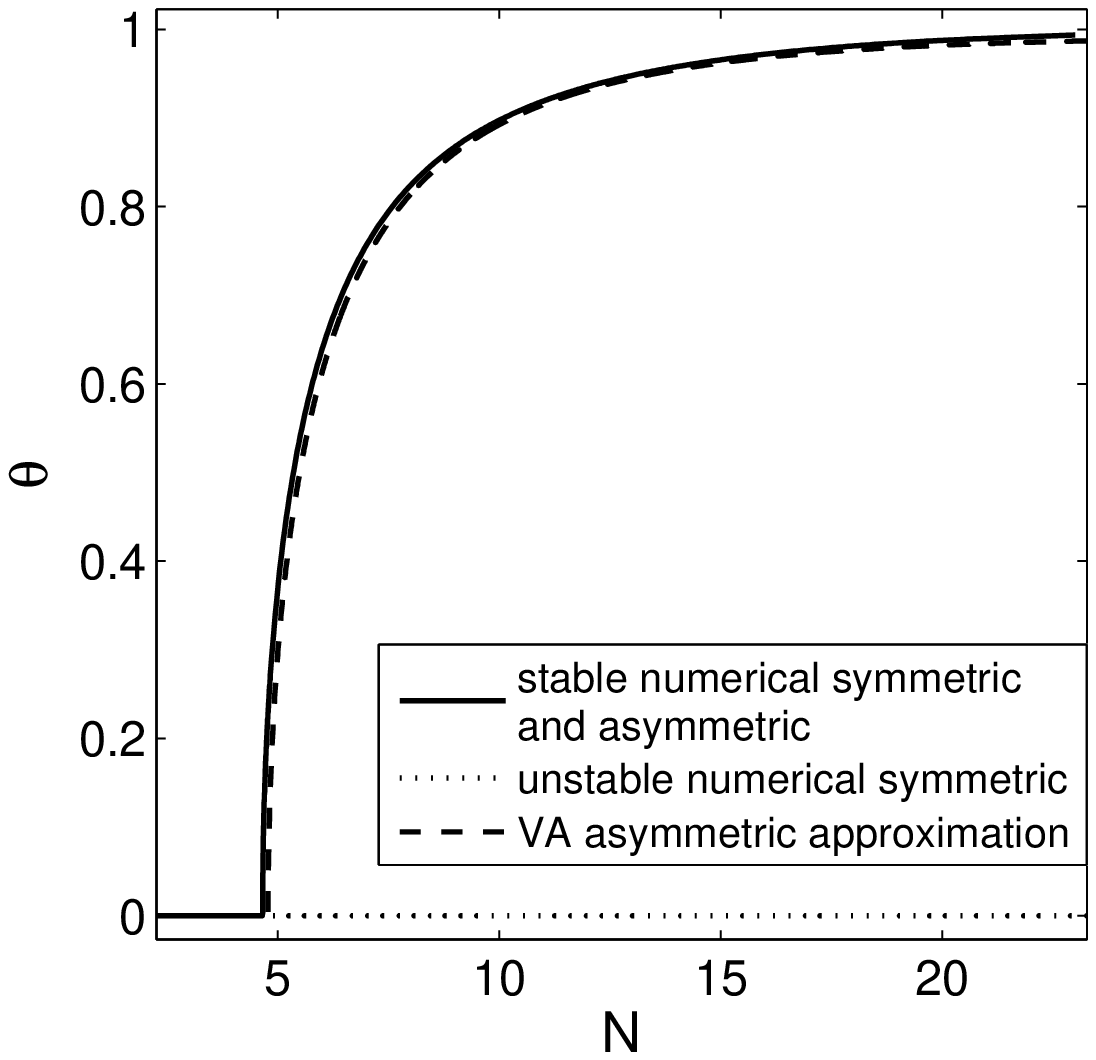}} \subfigure[]{%
\includegraphics[width=3.2in]{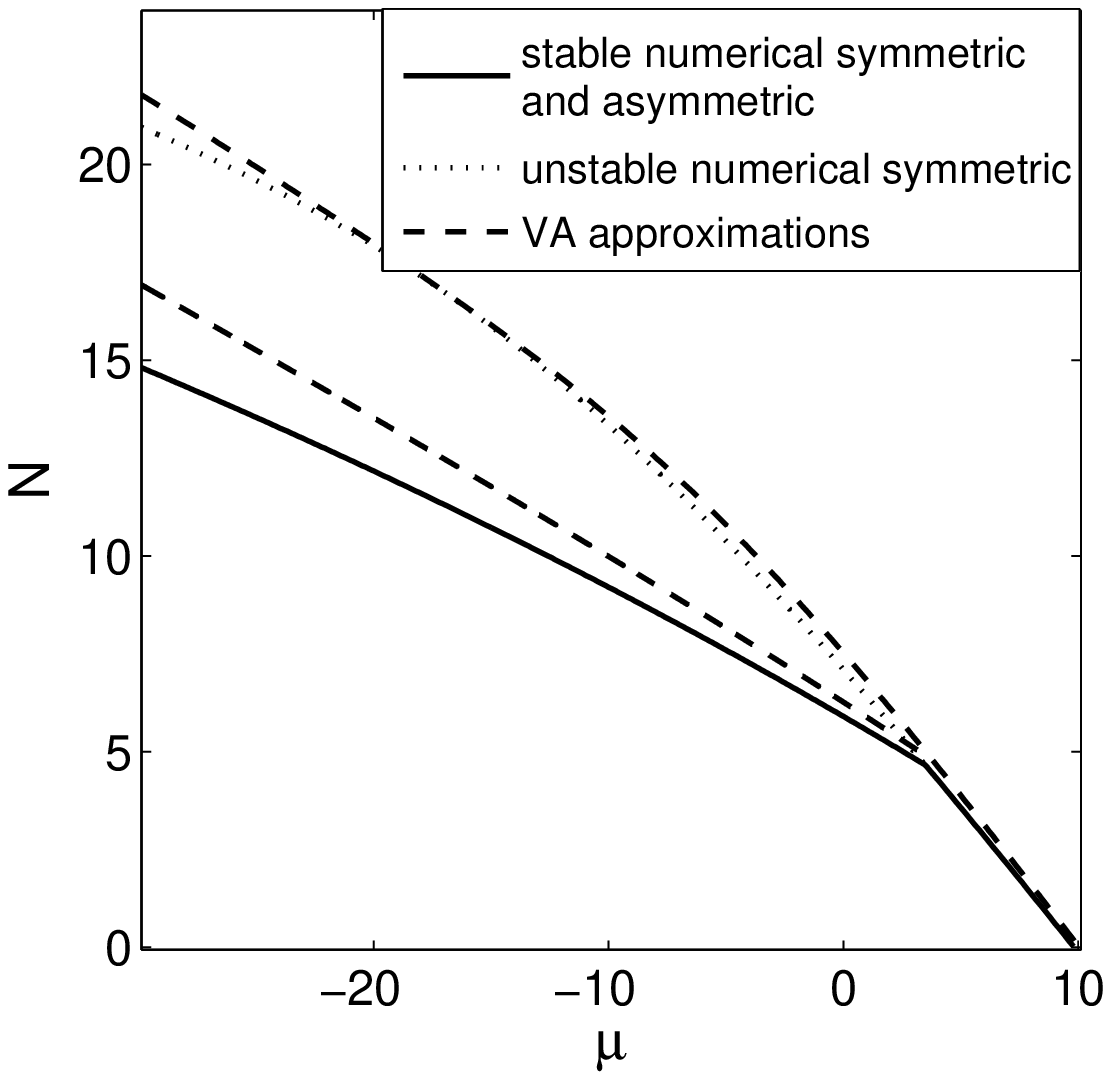}}
\caption{{}(a) The asymmetry of the stationary solutions, defined as per
Eqs. (\protect\ref{N}) and (\protect\ref{theta}), as a function of the total
norm, at $\protect\varepsilon =3$. For the VA solution, the asymmetry is
calculated through Eq. (\protect\ref{Theta}). (b) The dependence between the
total norm and energy eigenvalue for the same solutions.}
\label{fig4}
\end{figure}

For $\varepsilon $ which is not too small, both the numerical solution of
the VA equations (\ref{d/db}) and (\ref{dd/dadc}), and the numerical
solution of the stationary GPE (\ref{NLSE}) give rise to a characteristic
picture of the \textit{supercritical bifurcation} \cite{Iooss}, which is
displayed in Fig. \ref{fig4}(a), showing the asymmetry, defined as per Eqs. (%
\ref{N}) and (\ref{theta}), versus the total norm. In a broad range of
values of $\varepsilon $ (provided that it is not too small, see below),
this diagram, as predicted by the VA through Eq. (\ref{Theta}), is virtually
identical to its numerical counterpart.

The solution of the VA equations (\ref{d/db}) and (\ref{dd/dadc}) also
predicts $N(\mu )$ dependences for the symmetric and asymmetric solutions,
which are displayed and compared to their GPE-produced counterparts in Fig. %
\ref{fig4}(a). It is seen that the discrepancy is larger than in the
bifurcation diagram in Fig. \ref{fig4}(a), but, still, the VA provides
reasonable accuracy.

\begin{figure}[tbp]
\subfigure[]{\includegraphics[width=3.2in]{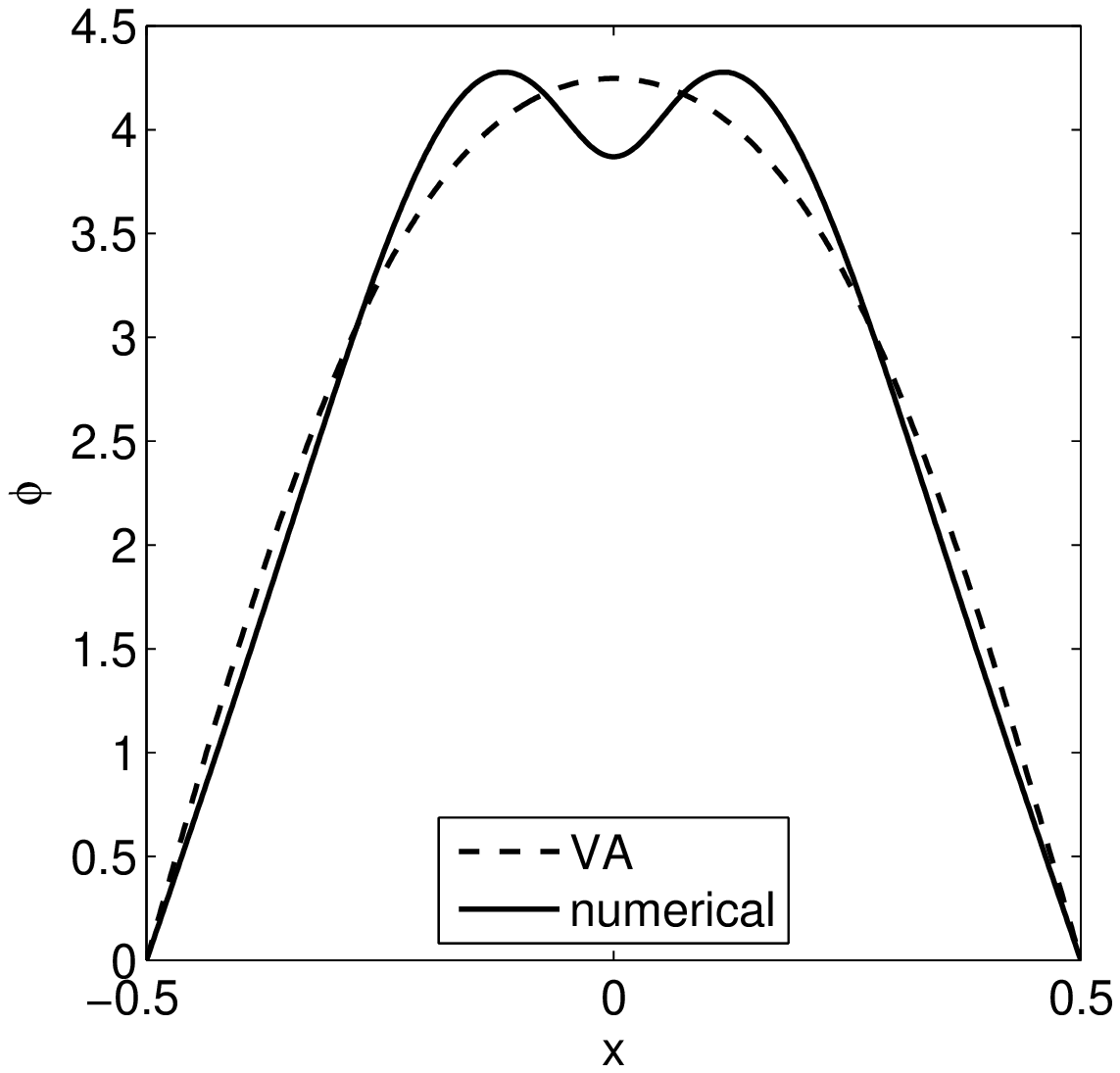}} \subfigure[]{%
\includegraphics[width=3.2in]{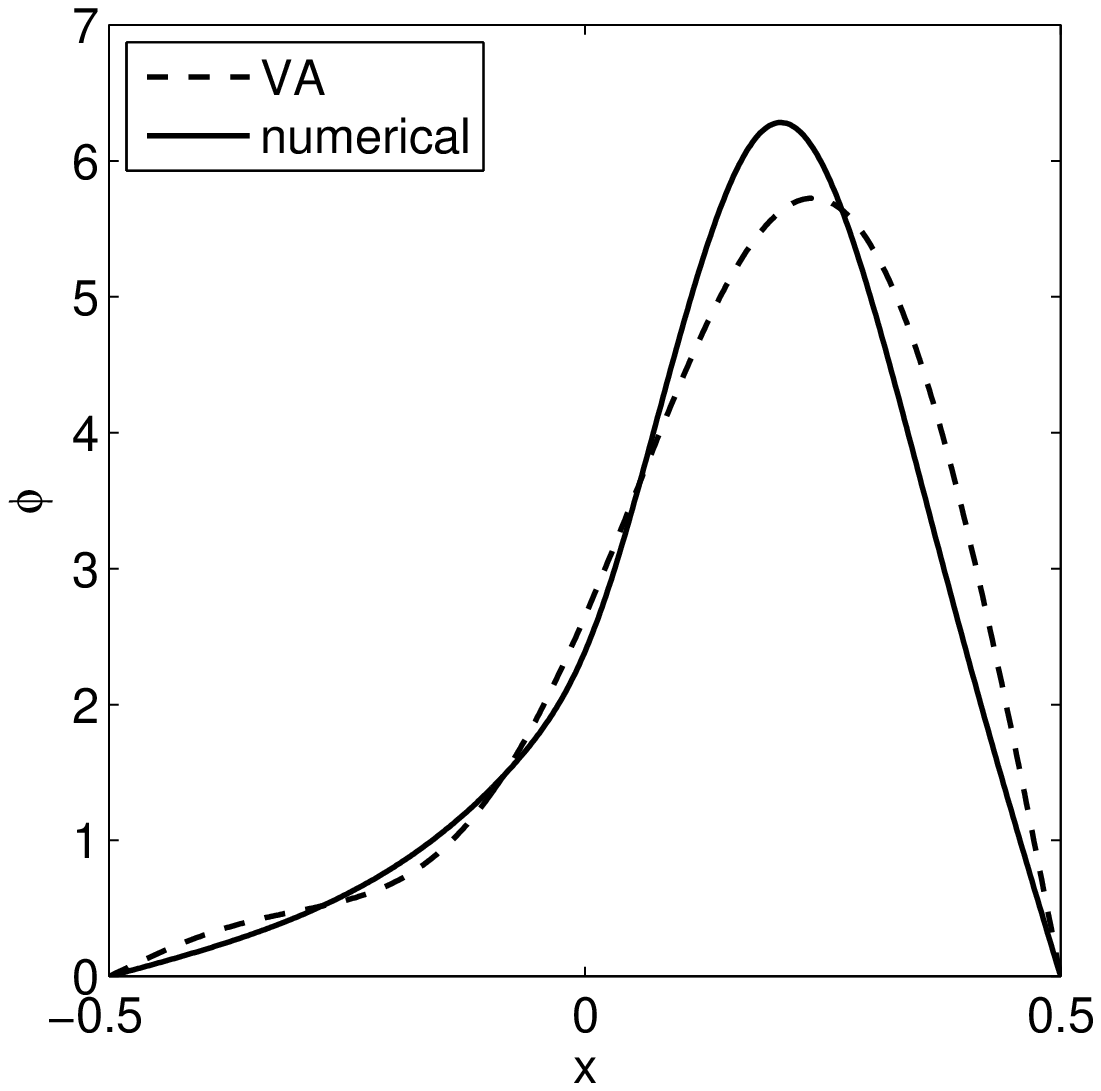}}
\caption{Typical examples of unstable symmetric (a) and stable asymmetric
(b) modes, produced by both the VA and numerical solution of Eq. (\protect
\ref{NLSE}) for $\protect\varepsilon =3$ and $N=10$. In panel (a), the
respective energy eigenvalues are $\protect\mu _{\mathrm{VA}}=-3.53$ and $%
\protect\mu _{\mathrm{num}}=-4.39$. In panel (b), they are $\protect\mu _{%
\mathrm{VA}}=-10.04$ and $\protect\mu _{\mathrm{num}}=-12.90$.}
\label{fig5}
\end{figure}

As concerns particular profiles of the symmetric and asymmetric modes,
produced by the VA and numerical solution of Eq. (\ref{NLSE}), typical
examples, displayed in Fig. \ref{fig5}, demonstrate reasonable agreement
between both. The remaining discrepancies between the variational and
numerical profiles can be further reduced if more spatial harmonics are
added to ansatz (\ref{ans}), at the cost of making the VA equations (\ref%
{d/db}) and (\ref{dd/dadc}) more cumbersome.

Another conclusion following from Fig. \ref{fig4}(b) is that the branches of
the $N(\mu )$ dependence satisfy the Vakhitov-Kolokolov criterion, $dN/d\mu
<0$, which is a necessary (but not sufficient) condition for stability of
localized modes supported by any self-attractive nonlinearity \cite{VK},
\cite{Berge,Kuzn}. As concerns the full stability, it was checked, as
mentioned above, by means of the numerical solution of the eigenvalue
problem based on Eq. (\ref{linearized}), and by direct simulations of Eq. (%
\ref{psi2}) as well, the conclusion being typical for the supercritical
bifurcation \cite{Iooss,book}: the symmetric mode at $N<N_{\mathrm{bif}}$
and the asymmetric ones at $N>N_{\mathrm{bif}}$ are stable, while the
symmetric branch is unstable at at $N>N_{\mathrm{bif}}$, when it coexists
with the asymmetric counterparts.

\begin{figure}[tbp]
\includegraphics[width=3.2in]{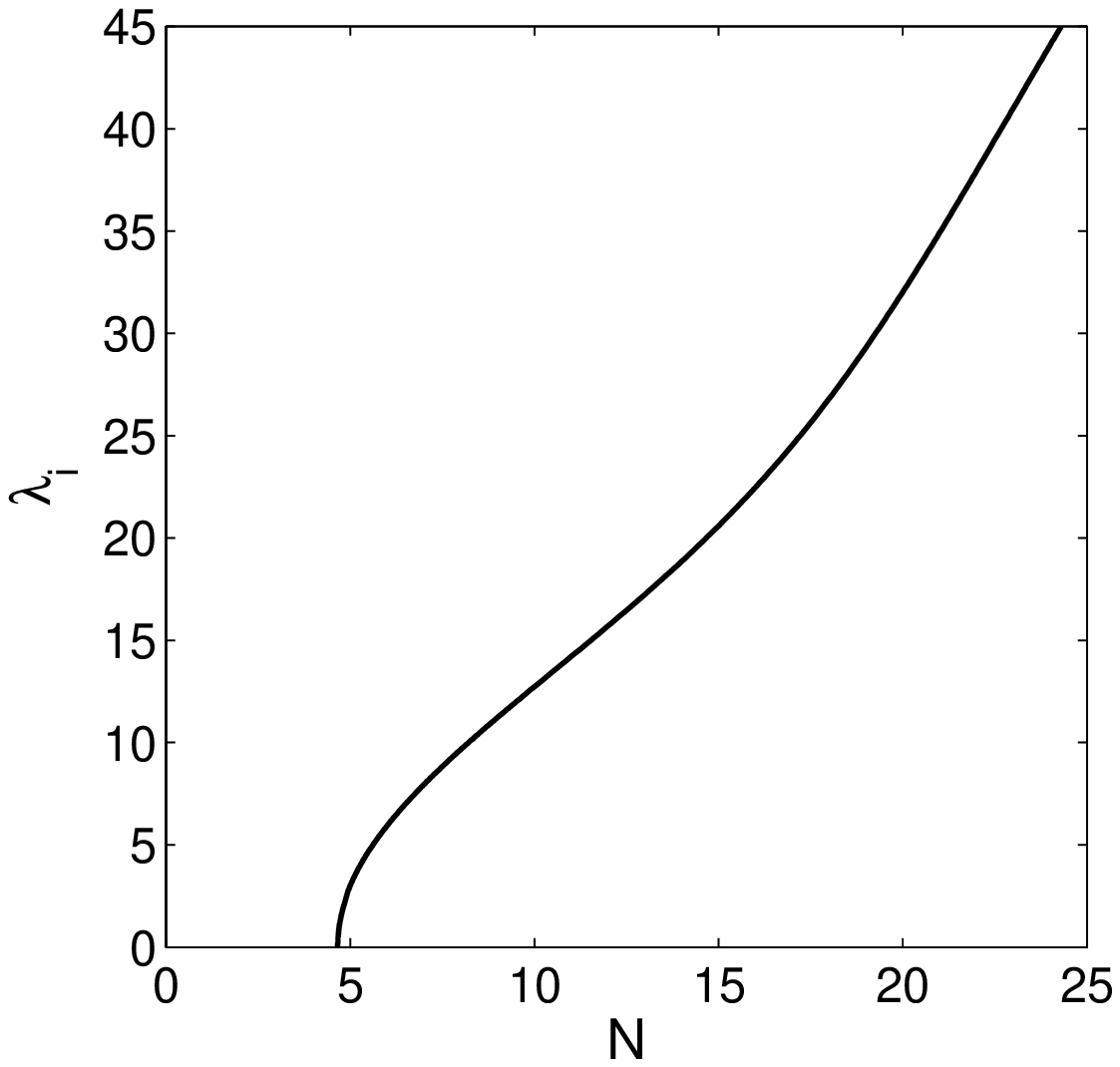}
\caption{{}The instability growth rate, $\protect\lambda _{\mathrm{i}}\equiv
\mathrm{Im}(\protect\lambda )$, of small perturbations added to the unstable
symmetric mode at $\protect\varepsilon =3$ and $N>N_{\mathrm{bif}}$, see Eq.
(\protect\ref{pert}). The growth rate was found as a numerical solution of
Eq. (\protect\ref{linearized}). The instability sets in at exactly the same
bifurcation point at which the SSB starts in Fig. \protect\ref{fig4}(a).}
\label{fig6}
\end{figure}

\begin{figure}[tbp]
\includegraphics[width=3.2in]{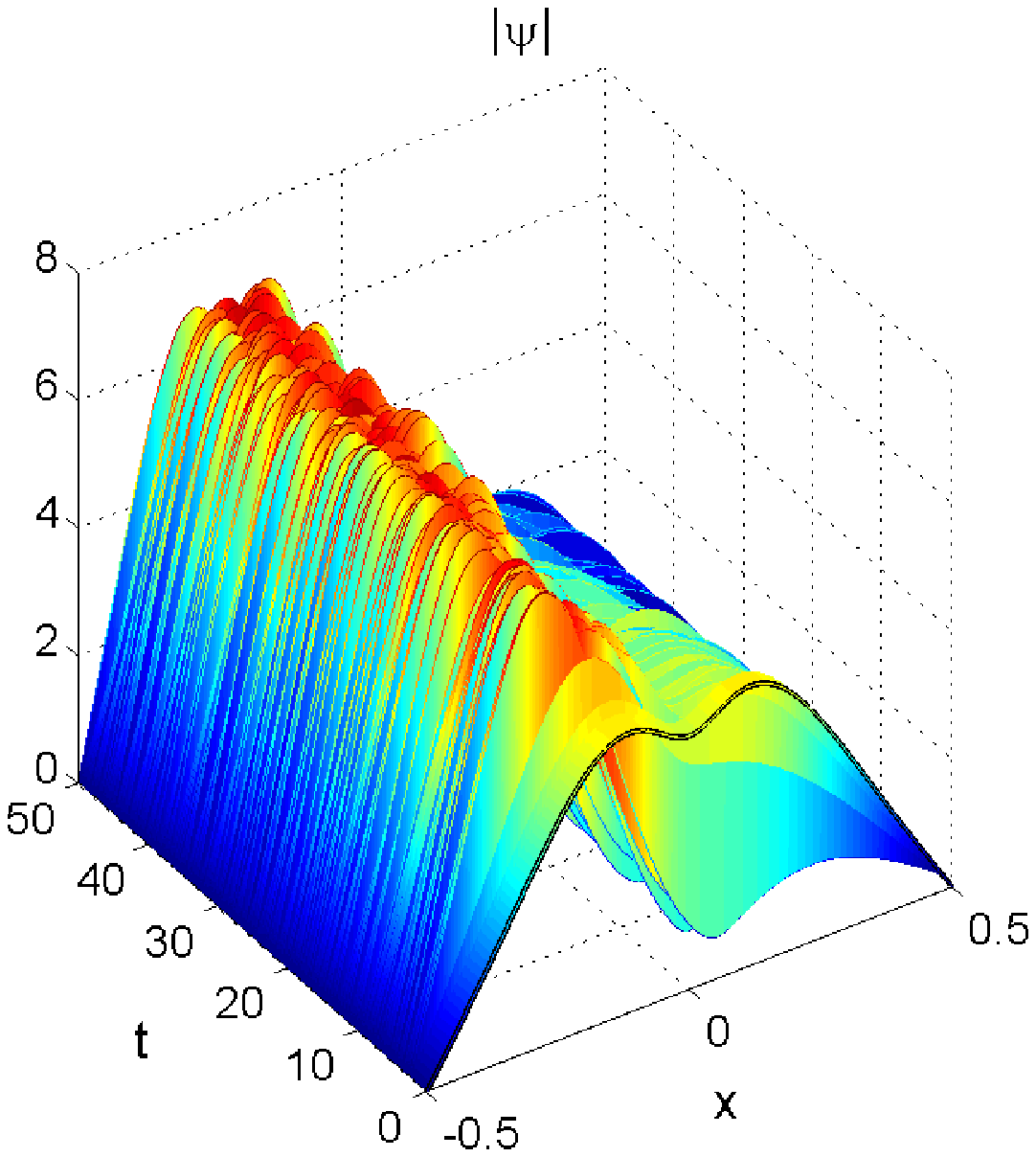}
\caption{{}A typical example of the spontaneous conversion of a perturbed
unstable symmetric mode into a stable asymmetric one, at $\protect%
\varepsilon =3$, $\protect\mu =3$, and $N=10.39$.}
\label{fig7}
\end{figure}

The instability of the symmetric branch at $N>N_{\mathrm{bif}}$ is
characterized, in Fig. \ref{fig6}, by the dependence of the respective
growth rate, $\mathrm{Im}(\lambda )$ [see Eq. (\ref{pert})], on $N$. A
typical example of the evolution of unstable symmetric modes is displayed in
Fig. \ref{fig7}, which clearly shows a trend to conversion of this unstable
mode into its stable asymmetric counterpart, with the same norm.

\subsection{The SSB at small $\protect\varepsilon $}

\begin{figure}[tbp]
\subfigure[]{\includegraphics[width=3.2in]{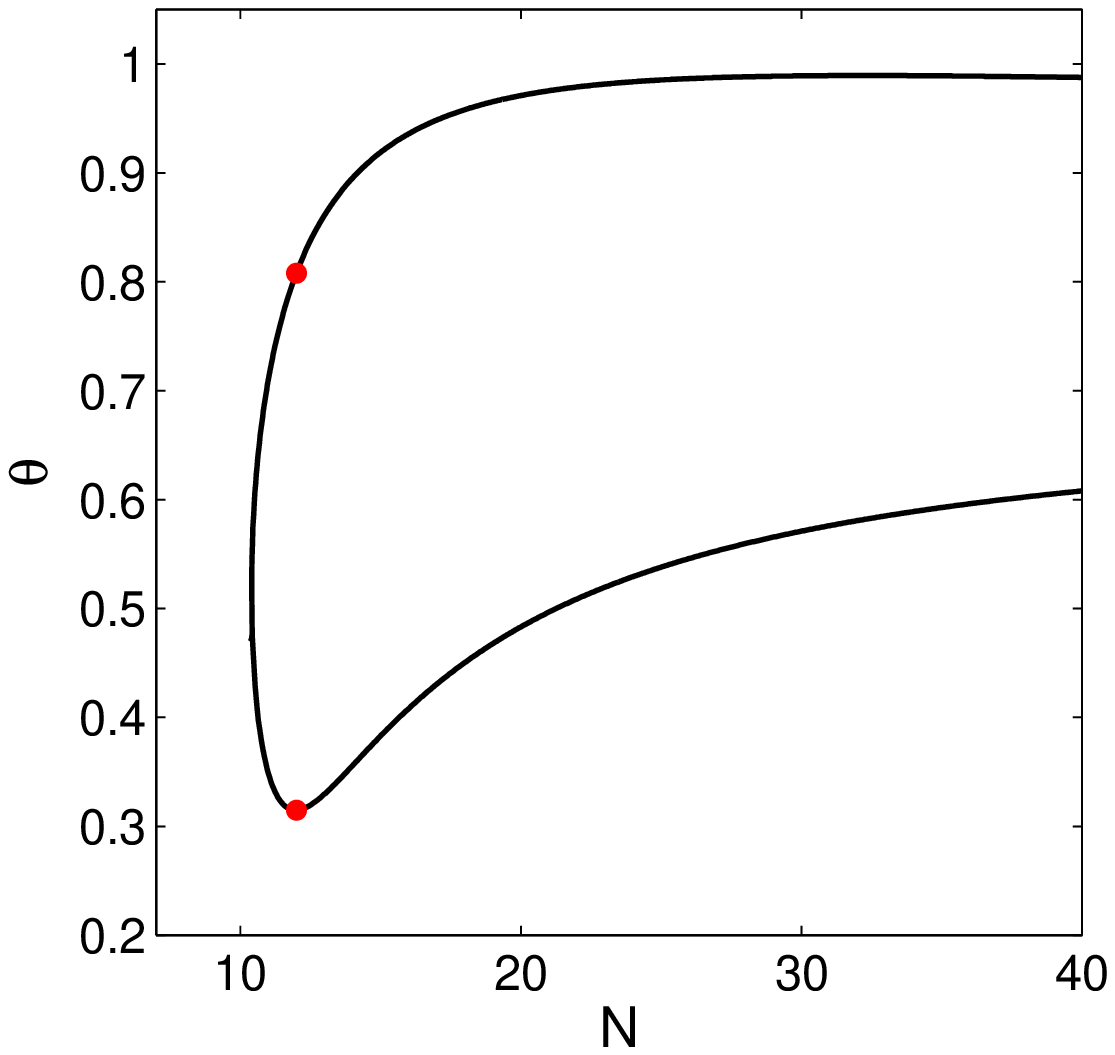}} \subfigure[]{%
\includegraphics[width=3.2in]{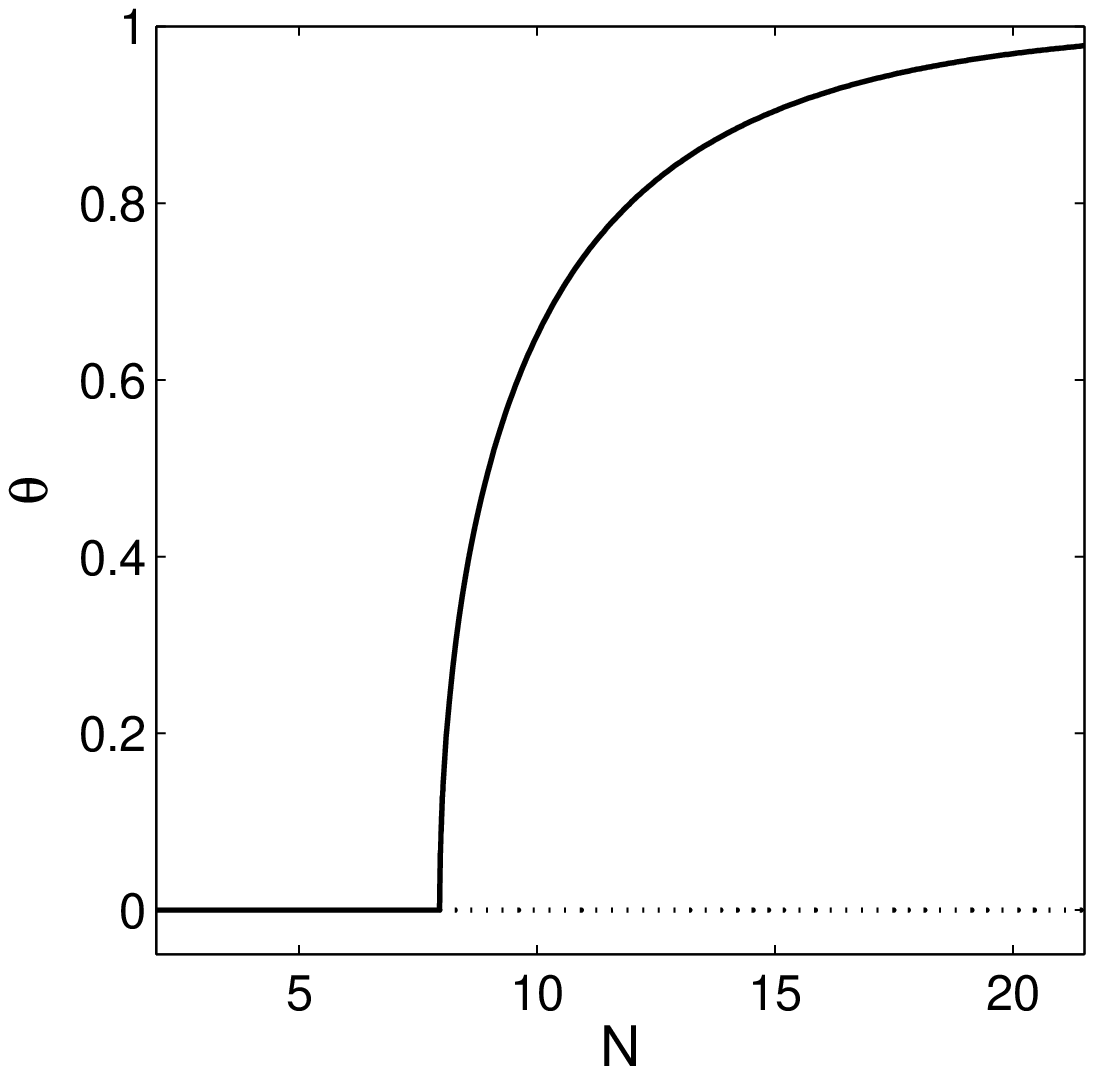}} \subfigure[]{%
\includegraphics[width=3.2in]{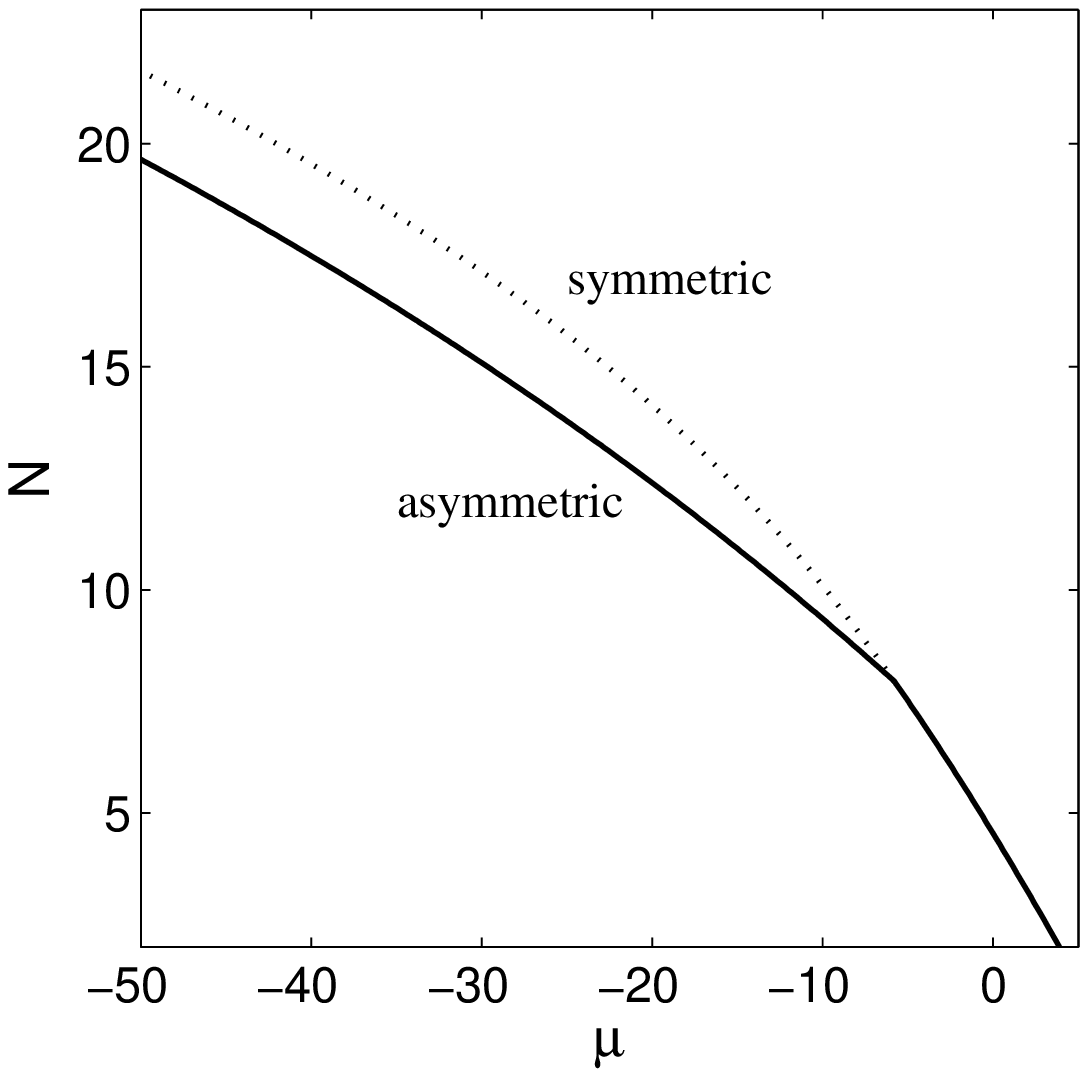}}
\caption{{}(a) The bifurcation diagram shown by means of the $\Theta (N)$
dependence, as predicted by the VA at $\protect\varepsilon =1$. Red dots
denote two asymmetric solutions for $N=12$, shown in Fig. \protect\ref{fig9}%
. (b) The counterpart of the bifurcation diagram from panel (a), as produced
by the numerical solution of Eq. (\protect\ref{NLSE}). (c) Numerically
generated $N(\protect\mu )$ dependences corresponding to panel (b).}
\label{fig8}
\end{figure}

The situation is different at smaller values of strength $\varepsilon $ of
the splitting barrier. Namely, the direct numerical solution demonstrates
that the SSB bifurcation keeps its supercritical character, while the VA
predicts a drastic change of the bifurcation diagram: detachment of the
asymmetric branch from the symmetric one, at $\varepsilon <\varepsilon _{%
\mathrm{cr}}^{\mathrm{(VA)}}\approx 1.50$, as shown in Fig. \ref{fig8} for $%
\varepsilon =1$. In fact, the comparison of Figs. \ref{fig8}(a) and (b)
demonstrates that the top branch of the $\theta (N)$ dependence is correctly
predicted by the VA, while the bottom branch, which formally demonstrates
the detachment of the asymmetric modes from the symmetric ones, is an
artifact.

\begin{figure}[tbp]
\includegraphics[width=3.2in]{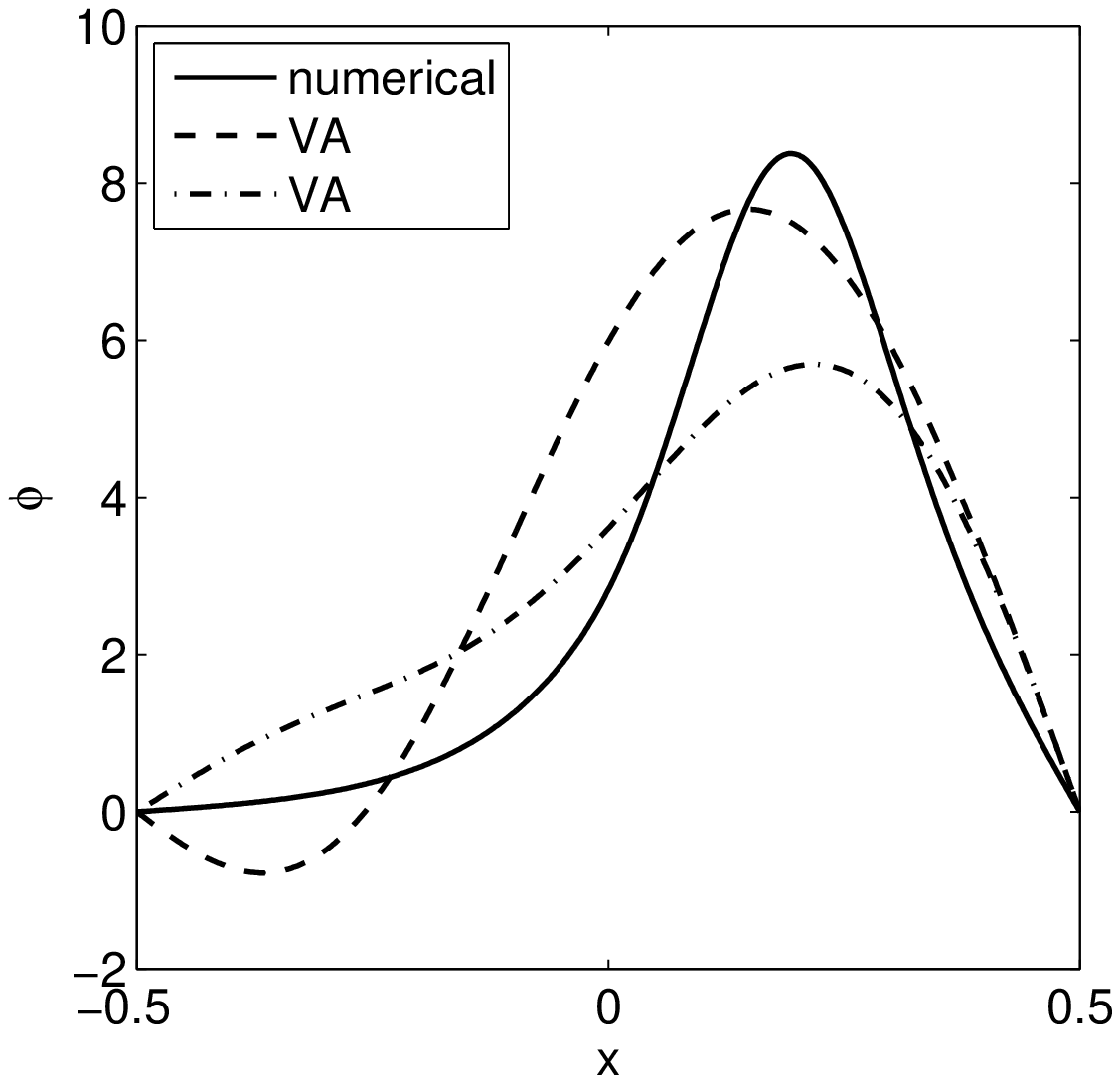}
\caption{{}Dashed and dashed-dotted profiles represent the VA-predicted
asymmetric solutions corresponding, respectively, to the top and bottom red
dots in Fig. \protect\ref{fig8}(a), obtained at $\protect\varepsilon =1$ and
$N=12$. The continuous profile depicts the solution obtained from the
numerical solution of Eq. (\protect\ref{NLSE}) with the same parameters. The
values of the energy eigenvalue pertaining to the dashed, dashed-dotted, and
continuous profiles are $\protect\mu =-24.7$, $-20.7$, and $-32.0$,
respectively.}
\label{fig9}
\end{figure}

The partly wrong shape of the bifurcation diagram predicted by the VA in the
case of $\varepsilon <\varepsilon _{\mathrm{cr}}^{\mathrm{(VA)}}$ is
explained by the inadequacy of ansatz (\ref{ans}) in this case. Indeed, the
comparison of typical shapes of the asymmetric solutions produced by the VA
with their numerically generated counterpart, shown in Fig. \ref{fig9},
demonstrates that the variational ansatz predicts the solutions belonging to
the top branch of the bifurcation diagram [see Fig. \ref{fig8}(a)] in a
qualitatively correct form, while the solutions belonging to the bottom
branch have no numerically found counterparts, being an artifact of the VA.
On the other hand, it is relevant to stress that the other analytical
approximation, based on soliton ansatz (\ref{sol}), (\ref{ghost}), correctly
demonstrates that the supercritical SSB bifurcation occurs at small $%
\varepsilon $ too.

\subsection{Excited modes}

As said above, the shape of the first (spatially antisymmetric, alias odd)
excited mode in the model based on Eq. (\ref{psi}), with the ideal $\delta $%
-functional splitting barrier, is not affected by the barrier. However,
stability of the antisymmetric mode against antisymmetry-breaking
perturbations may be altered in the presence of the barrier. We have
addressed this issue by means of the numerical analysis, replacing the ideal
$\delta $-function by its regularized counterpart (\ref{delta}).

\begin{figure}[tbp]
\subfigure[]{\includegraphics[width=3.2in]{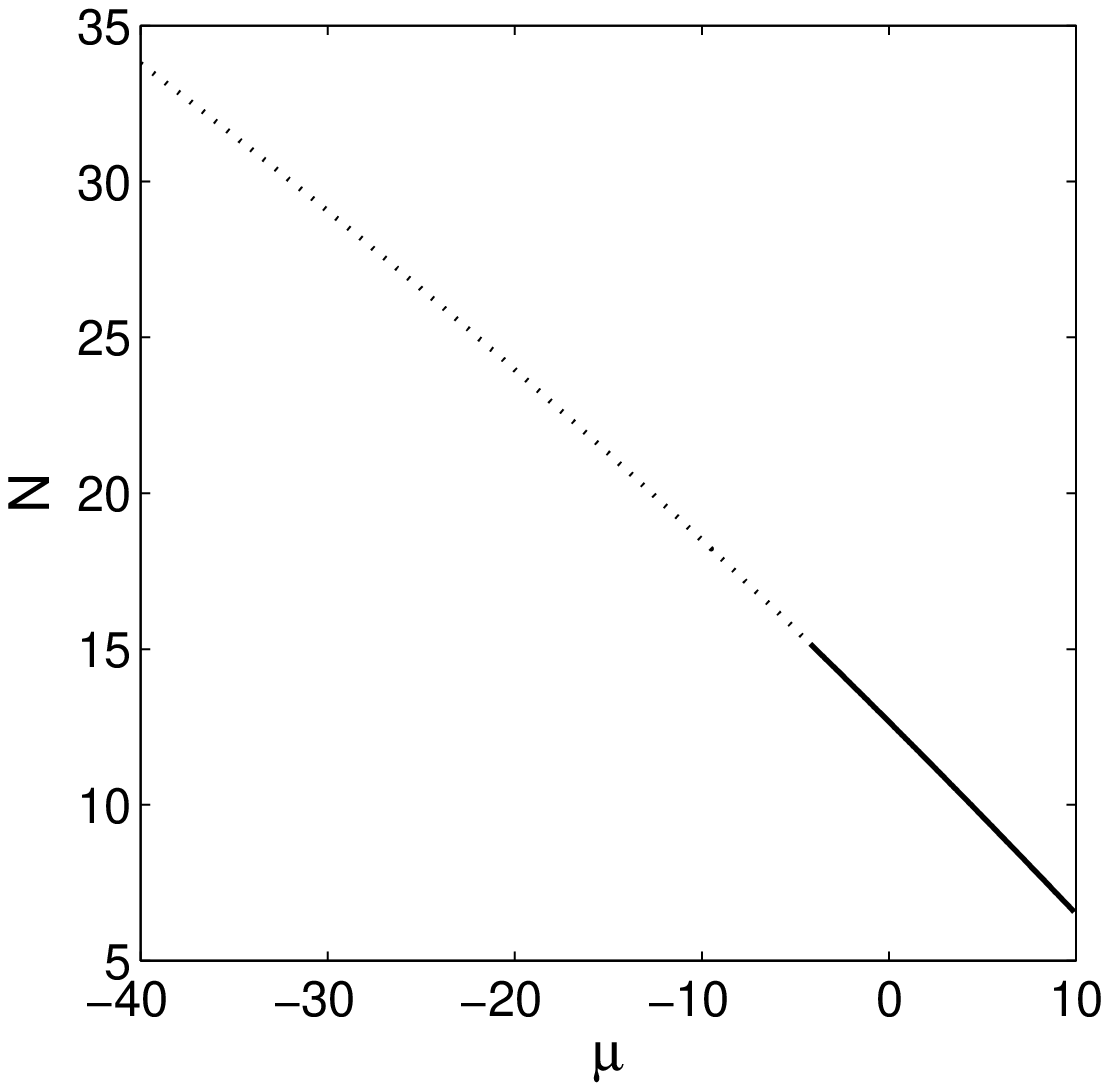}} \subfigure[]{%
\includegraphics[width=3.2in]{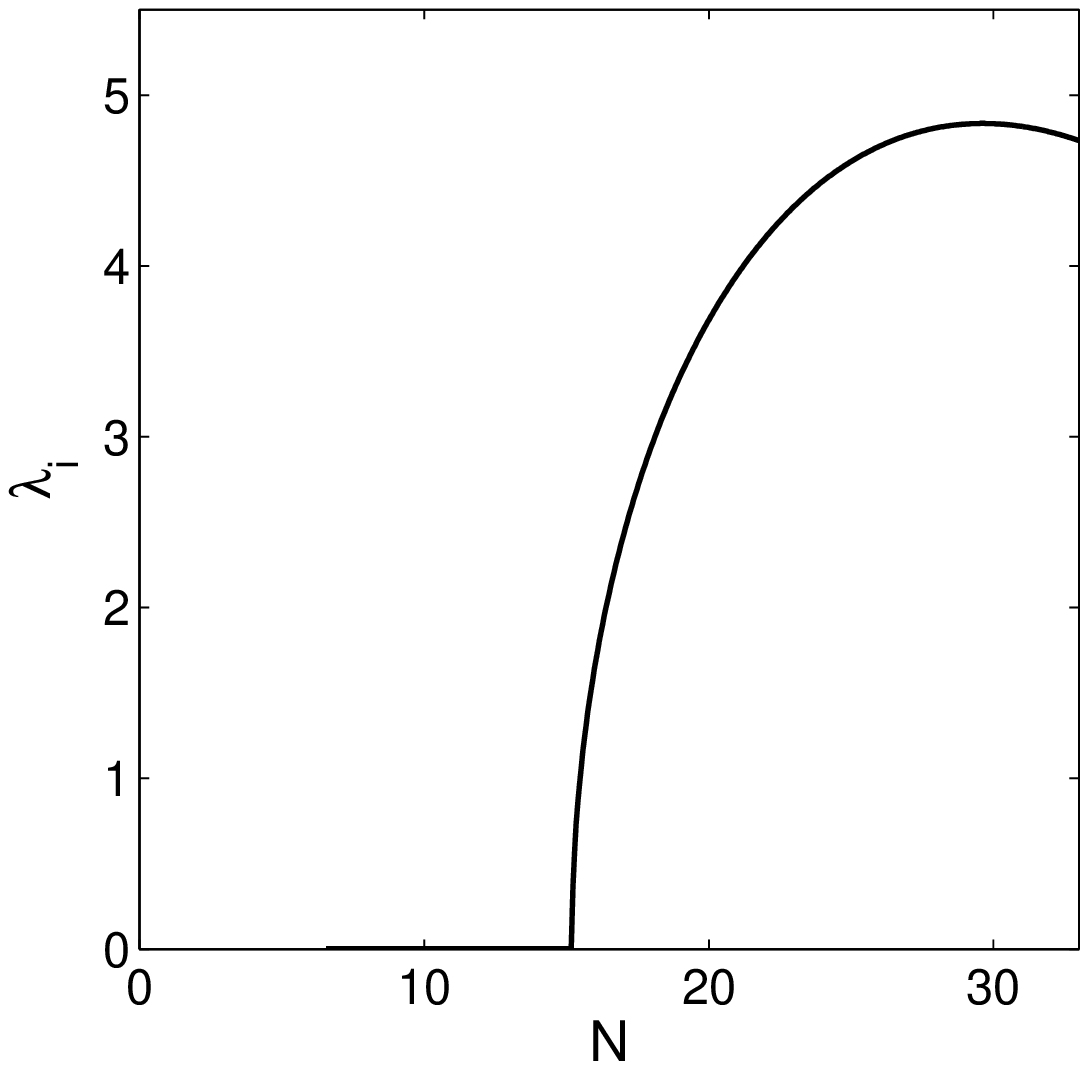}}
\caption{{}(a) The numerically generated $N(\protect\mu )$ dependence for
the family of odd (antisymmetric) modes at $\protect\varepsilon =3$. Stable
and unstable subfamilies are represented by the continuous and dotted
segments, respectively. (b) The instability growth rate [see Eq. (\protect
\ref{pert})] for the unstable portion of the family versus $N$.}
\label{fig10}
\end{figure}

\begin{figure}[tbp]
\includegraphics[width=3.2in]{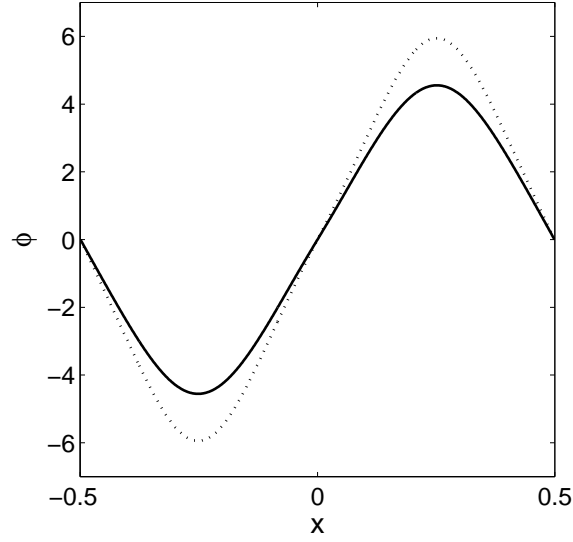}
\caption{{}Examples of stable and unstable (shown by the continuous and
dotted lines, respectively) odd modes, for $\protect\varepsilon =3$. The
respective values of the norm and energy eigenvalue are $N=9.63$, $\protect%
\mu =+5$ and $N=15.63,\protect\mu =-5$.}
\label{fig11}
\end{figure}

\begin{figure}[tbp]
\includegraphics[width=3.2in]{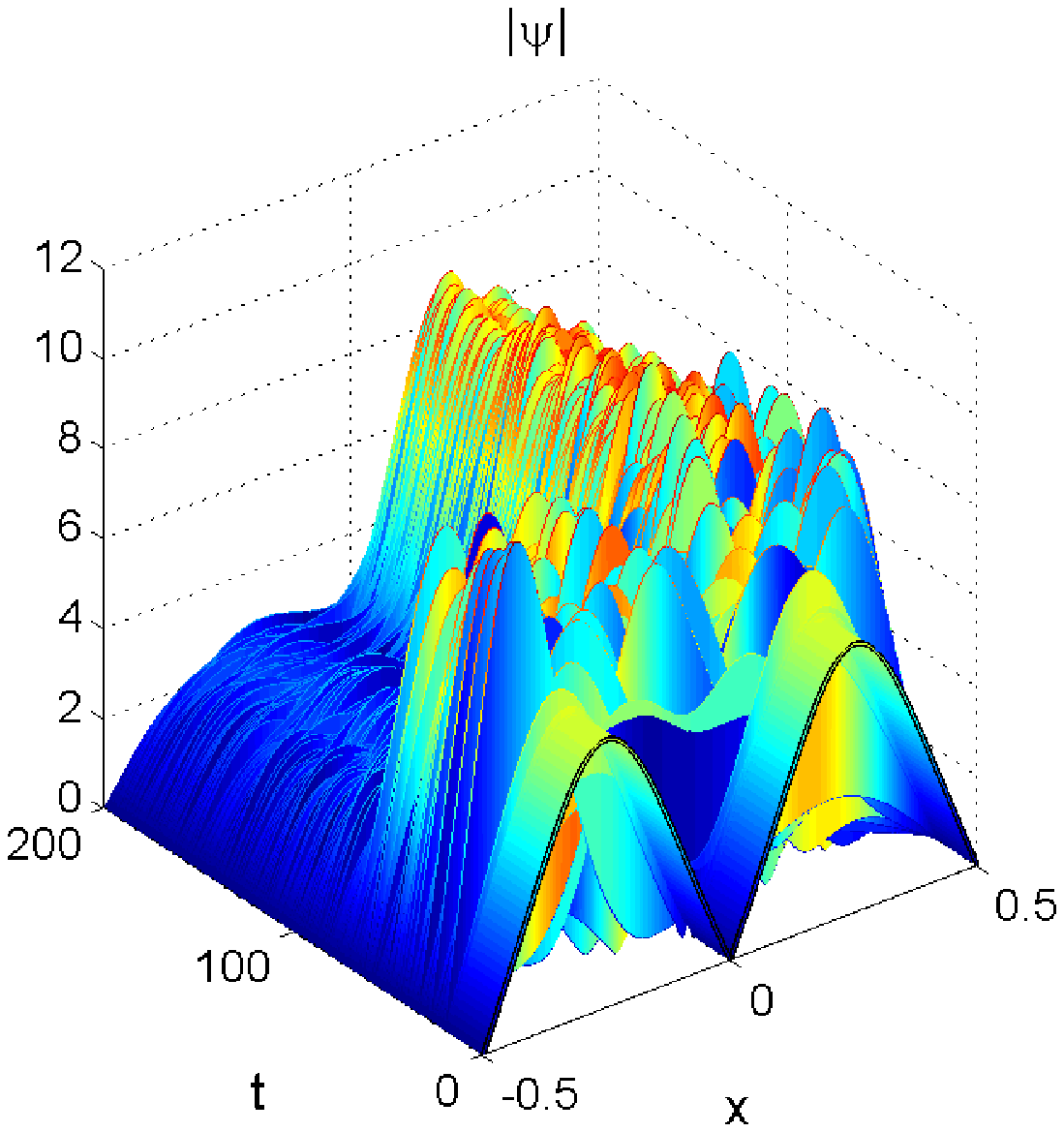}
\caption{{}(Color online) An example of the spontaneous evolution of an
unstable odd mode, shown by the dotted line in Fig. \protect\ref{fig11} (for
$\protect\varepsilon =3,$ $N=15.63$, $\protect\mu =-5$), towards a stable
asymmetric state.}
\label{fig12}
\end{figure}

As shown in Fig. \ref{fig10}, it was found that the odd\ modes are stable if
their norm is small enough, and become unstable when $N$ exceeds a certain
critical value; however, the instability is oscillatory, not being related
to any antisymmetry-breaking bifurcation. Examples of stable and unstable
odd modes are displayed in Fig. \ref{fig11}, with a typical scenario of the
evolution of an unstable one presented in Fig. \ref{fig12}. It is observed
that unstable odd modes tend to spontaneously transform themselves into
stable asymmetric modes existing at the same value of $N$.

\begin{figure}[tbp]
\includegraphics[width=3.2in]{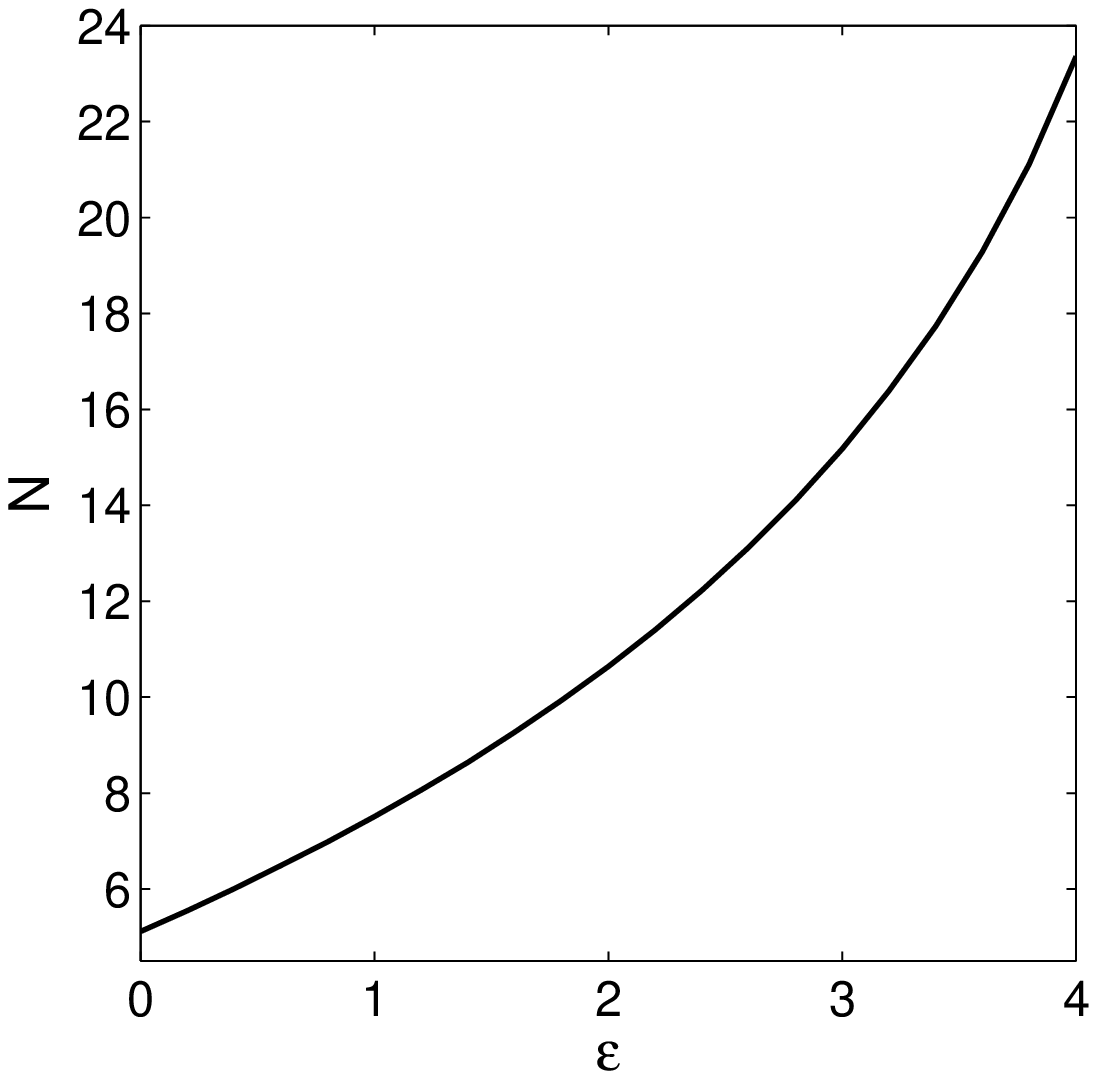}
\caption{{}Antisymmetric (odd) modes, which are the lowest excited states,
are stable in the region below the boundary shown here in the plane of $%
\left( \protect\varepsilon ,N\right) $.}
\label{fig13}
\end{figure}

Results of the systematic analysis of the odd modes are collected in Fig. %
\ref{fig13}, which shows their instability threshold (in terms of the norm)
versus $\varepsilon $. It is seen that the stability interval of the odd
modes is smallest ($N<N_{\mathrm{cr}}\approx 5.12$) in the absence of the
splitting barrier, i.e., at $\varepsilon =0$. The increase of $\varepsilon $
leads to expansion of the stability range, which may be explained by dint of
the following argument. In the case when both $\varepsilon $ and $N$ are
large, the odd mode may be considered as a superposition of two narrow
half-solitons (\ref{sol}), with norms $N\rightarrow N/2$, opposite signs,
and centers located at opposite points, $\pm \xi $. Because the energy of
the free soliton, determined by Eq. (\ref{H}), is $-(1/3)(N/2)^{3}$, the
instability of the odd state against the merger of the two half-solitons
into a single one with norm $N$, placed in either half-box (i.e., the
instability against the breaking of the antisymmetry), is driven by the
respective energy difference, $\Delta E=N^{3}/4$. On the other hand, Eq. (%
\ref{barrier}) suggests that, to pass the separating hurdle on the way to
the merger, a half-soliton must overcome an energy barrier of height $U_{%
\mathrm{barrier}}\sim \varepsilon N^{2}$. Thus, the value of the norm at the
instability threshold may be estimated, from condition $\Delta E\sim U_{%
\mathrm{barrier}}$, as $N_{\mathrm{cr}}\sim \varepsilon $. Eventually, in
the in the limit of $\varepsilon \rightarrow \infty $, the infinitely tall
barrier splits the potential box into two isolated ones, each lobe of the
odd mode carrying over into a stable GS (ground state) of the half-box,
which implies that $N_{\mathrm{cr}}(\varepsilon )\rightarrow \infty $ at $%
\varepsilon \rightarrow \infty $.

\begin{figure}[tbp]
\subfigure[]{\includegraphics[width=3.2in]{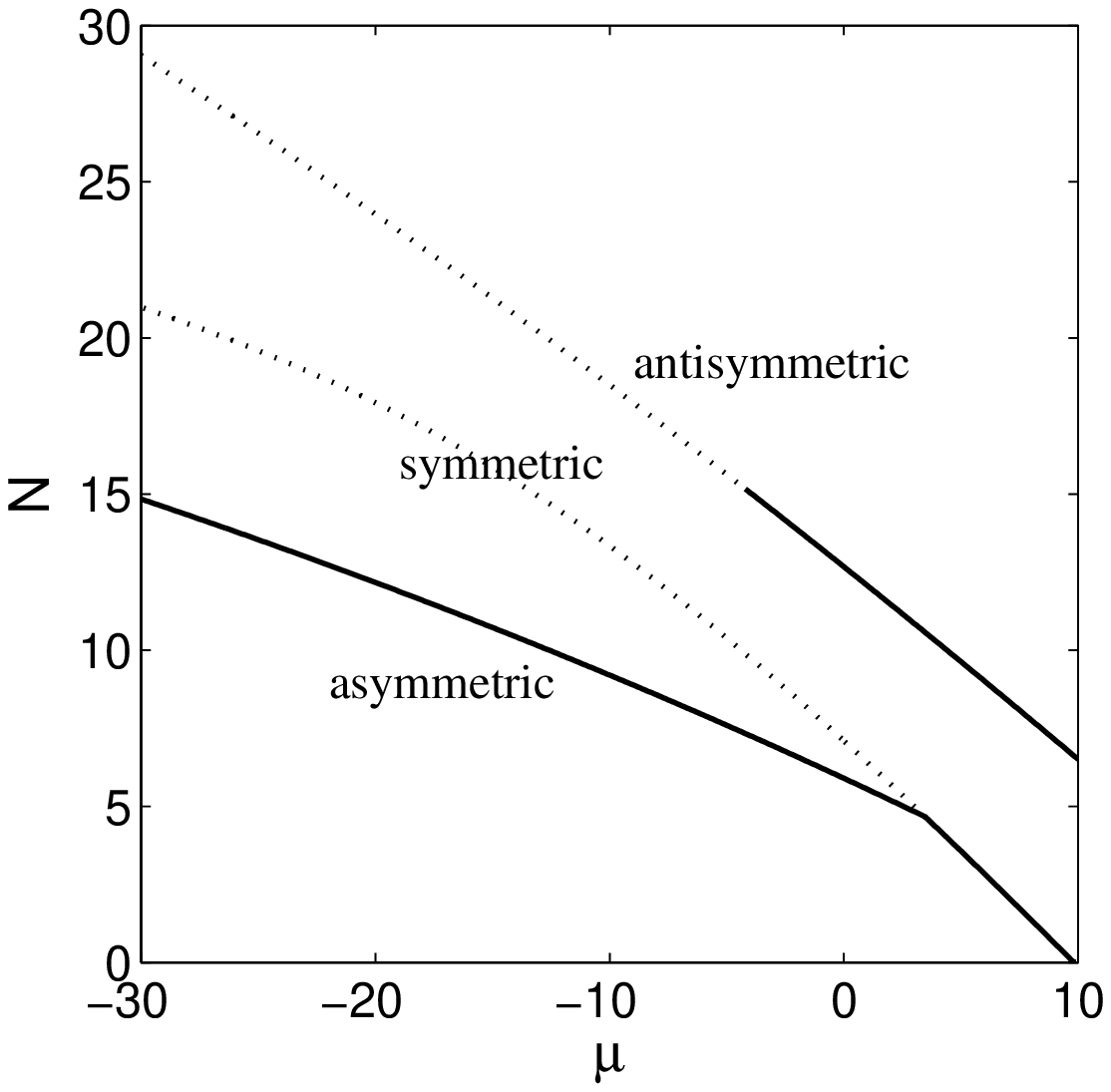}} \subfigure[]{%
\includegraphics[width=3.2in]{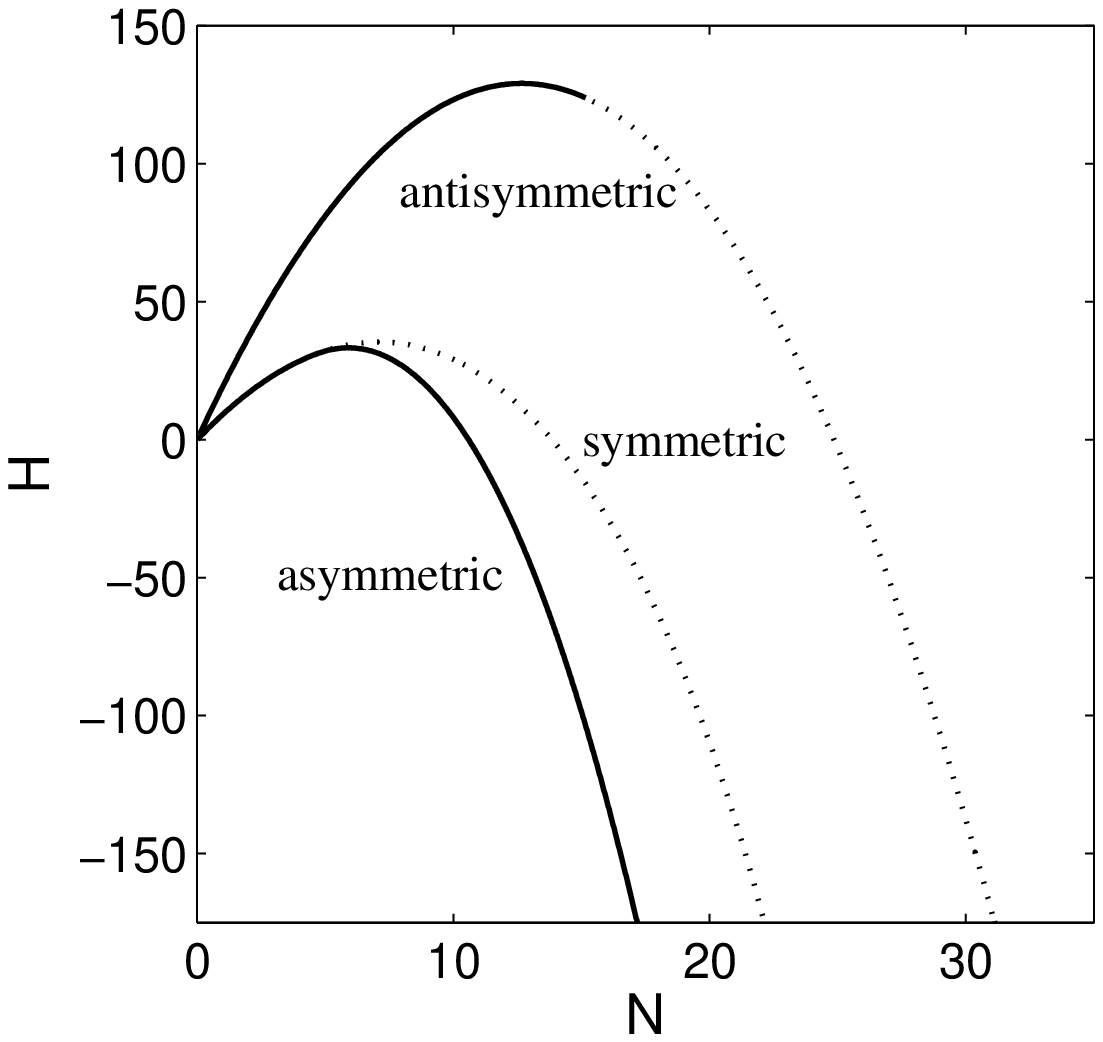}}
\caption{{}(a) The juxtaposed $N(\protect\mu )$ dependences for the
symmetric (even), antisymmetric (odd), and asymmetric modes at $\protect%
\varepsilon =3$. (b) The Hamiltonian versus the norm for the same modes,
calculated as per Eqs. (\protect\ref{H}) and (\protect\ref{tilde}). In both
panels, continuous and dotted segments represent stable and unstable
solutions, respectively.}
\label{fig14}
\end{figure}

\begin{figure}[tbp]
\subfigure[]{\includegraphics[width=3.2in]{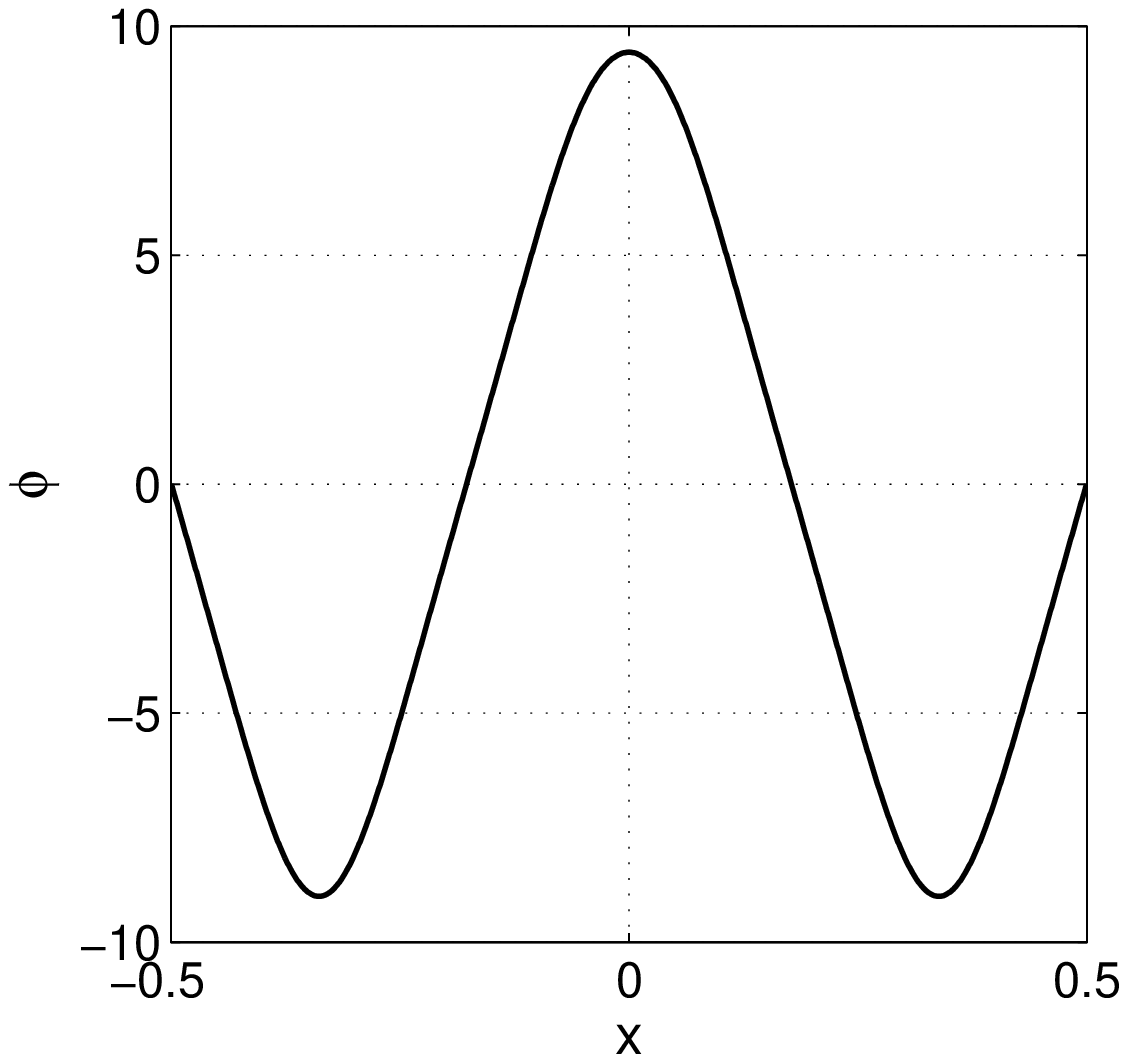}} \subfigure[]{%
\includegraphics[width=3.2in]{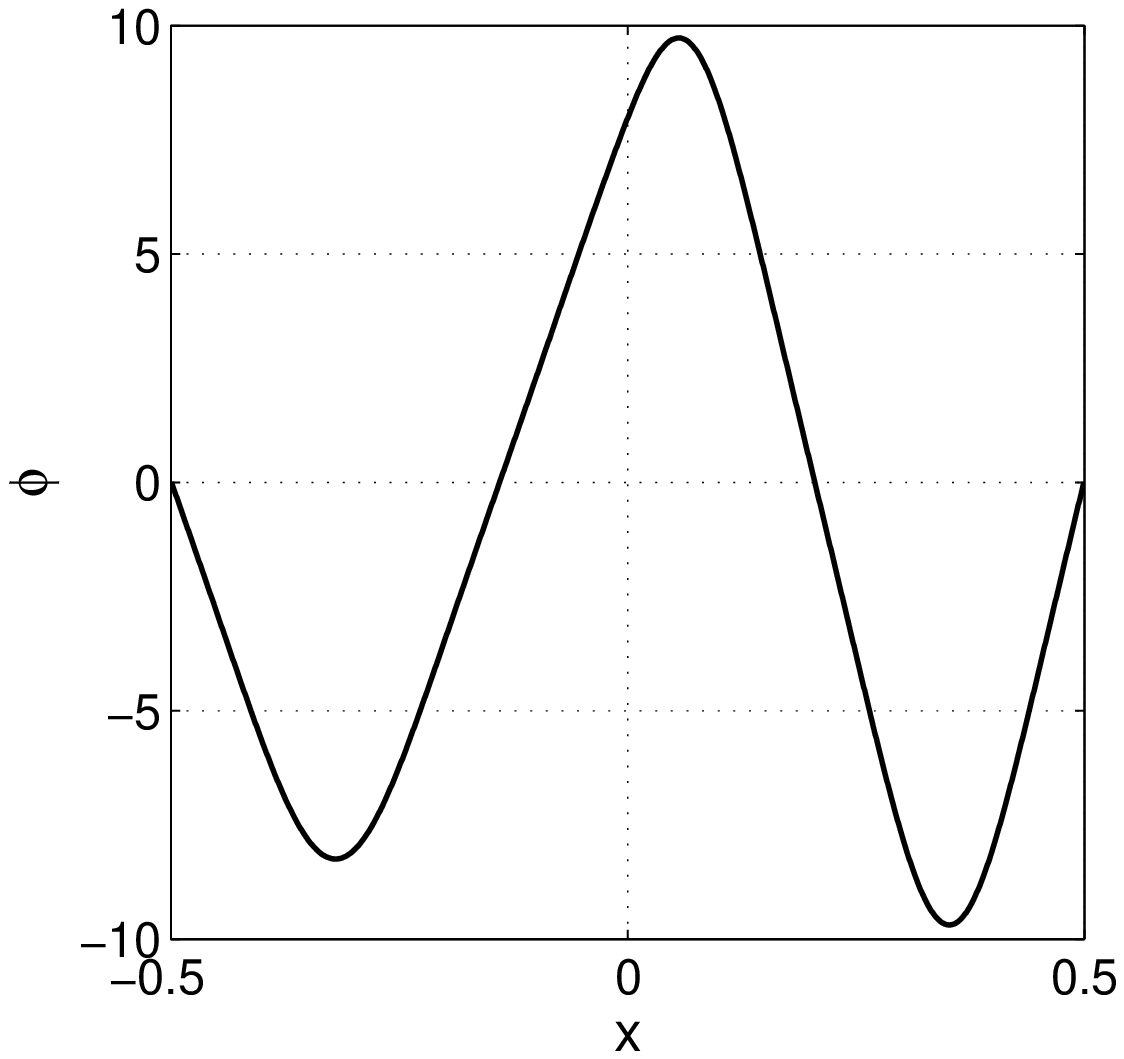}} \subfigure[]{%
\includegraphics[width=3.2in]{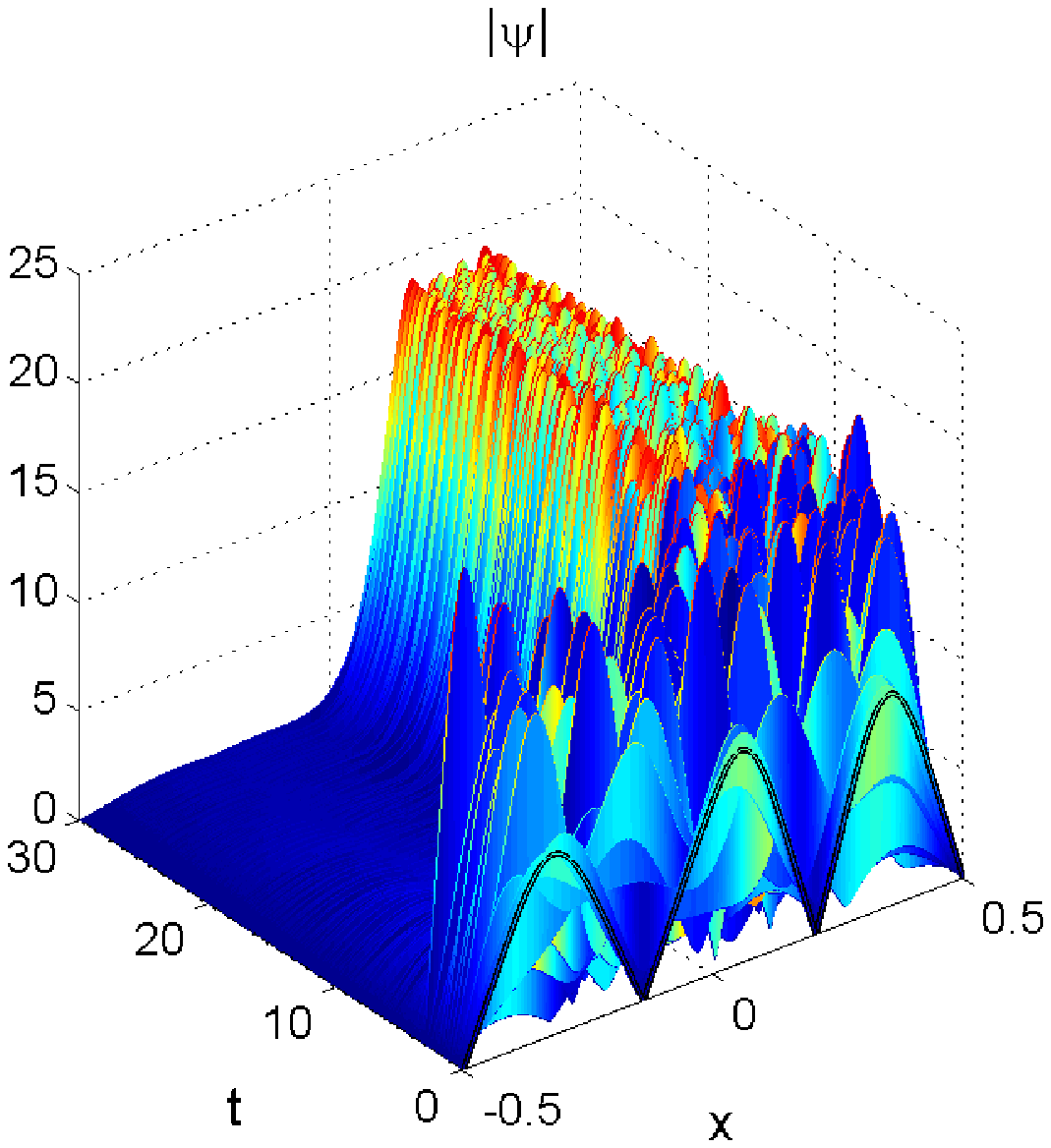}}
\caption{{}Numerically found second excited states, as solutions of Eq. (%
\protect\ref{NLSE}), for $\protect\varepsilon =3$ and $\protect\mu =-10$:
(a) a symmetric one, with norm $N=38.13$; (b) an asymmetric mode, with $%
N=37.39$. (c) Numerically simulated evolution of the unstable asymmetric
mode from (b). }
\label{fig15}
\end{figure}

Because the odd modes are stable at sufficiently small values of $N$, it
makes sense to compare their stability region with that of their symmetric
and asymmetric counterparts. Figure \ref{fig14}(a) presents the comparison
for $\varepsilon =3$. Further, to identify the system's GS, Fig. \ref{fig14}%
(b) displays the comparison of their Hamiltonian values as functions of the
norm. The Hamiltonian is calculated according to Eq. (\ref{H}), in which the
last term is replaced as per Eq. (\ref{delta}):%
\begin{equation}
\varepsilon \left\vert \psi (x=0)\right\vert ^{2}\rightarrow \varepsilon
\int_{-1/2}^{+1/2}\tilde{\delta}(x)\left\vert \psi (x)\right\vert ^{2}dx.
\label{tilde}
\end{equation}%
It is clearly seen that the odd mode is always an excited state, whose
Hamiltonian exceeds that of the coexisting (for the same $N$) stable
symmetric or asymmetric mode. Thus, the symmetric state, when it is stable,
and the asymmetric one, when it exists, represent the GS.

Lastly, we have also carried out a brief analysis for the second excited
state, i.e., the first even state existing above the GS. As well as the GS,
this mode undergoes the SSB, which gives rise, above the respective
bifurcation point, to coexisting symmetric and asymmetric versions of the
second excited state, see a typical example in Figs. \ref{fig14}(a,b).
However, a drastic difference from the GS is that not only the formally
existing symmetric mode, but also the coexisting asymmetric one, produced by
the SSB, is unstable, as shown in Figs. \ref{fig14}(c). As a result of its
evolution, the unstable asymmetric second-order state spontaneously evolves
towards a stable asymmetric GS which exists at the same norm.

\section{Conclusion}

The objective of this work is to carry out the systematic analysis of the
basic\ one-dimensional model which is capable to grasp the SSB (spontaneous
symmetry breaking) phenomenology. The model, which may be realized in BEC\
and nonlinear optics alike, is built as the infinitely deep potential box,
which is split into two wells by a narrow (delta-functional) barrier set at
the center. The barrier's strength, $\varepsilon $, is the single free
parameter in the scaled version of the model. The SSB is predicted in it by
means of two analytical approximations, which are valid in two limit cases,
\textit{viz}.,\ for strong or weak splitting of the potential box by the
central barrier. Another semi-analytical approach, based on the VA
(variational approximation), has been developed in the generic case.
Predictions of the analytical approximations have been verified by means of
comparison with systematically generated numerical results. It is inferred
that the system always gives rise to the supercritical SSB bifurcation of
the GS (ground state). The VA accurately predicts this finding at moderate
values of $\varepsilon $, but fails to do it at small $\varepsilon $, due to
the limited applicability of the underlying \textit{ansatz}. However, the
other analytical approximation for small $\varepsilon $, which is based on
the soliton ansatz, correctly describes that case. In addition to the GS,
the stability of the first and second excited states was investigated too.
The former one (a spatially odd mode) is destabilized at a critical value of
the norm. The second-order excited state, as well as the GS, features the
SSB bifurcation, but, unlike the GS, the asymmetric mode produced by this
bifurcation is unstable. In direct simulations, all unstable modes tend to
rearrange themselves into the symmetry-broken GS with the same norm.

As an extension of the work, it may be interesting to consider its
two-dimensional version, for a square-shaped infinitely deep potential box,
split by appropriate inner barriers, a new factor appearing in two
dimensions being a possibility of the collapse of the trapped modes. The
analysis of a two-component version of the system may be relevant too.

This work was supported, in part, by the joint program in physics between
the National Science Foundation (US) and Binational Science Foundation
(US-Israel), through grant No. 2015616.

\end{document}